\UseRawInputEncoding
\documentclass[a4paper,aps,prd,10pt,preprintnumbers,showpacs,twocolumn,superscriptaddress,nofootinbib,amsmath,amssymb]{revtex4-1}
\usepackage{graphicx}
\usepackage{cmap}
\usepackage[utf8]{inputenc}
\usepackage[T1]{fontenc}

\def\imo{i}
\def\re#1{Re(#1)}
\def\im#1{Im(#1)}
\def\K{{\cal K}}
\def\Order#1{{\cal O}\left(#1\right)}

\begin{document}

\title{Scalar, electromagnetic, and Dirac perturbations of regular black holes constituting  primordial dark matter}
\author{Bekir Can L{\"u}tf{\"u}o{\u g}lu}
\email{bekir.lutfuoglu@uhk.cz}
\affiliation{Department of Physics, Faculty of Science, University of Hradec Králové, Rokitanského 62/26, 500 03 Hradec Králové, Czech Republic}

\begin{abstract}
We study massless scalar, electromagnetic, and Dirac perturbations of the exact asymptotically flat regular black hole supported by a phantom Dirac--Born--Infeld scalar. Using the Pad\'e-improved WKB method, with a time-domain Prony check for the scalar fundamental mode, we compute representative quasinormal frequencies and find that larger regularity shifts the spectrum toward smaller oscillation frequencies and damping rates, whereas the quality factor changes only weakly. The spectral shifts remain well above the estimated numerical uncertainty, demonstrating that the DBI regularity scale leaves a robust spin-dependent imprint on ringdown.
\end{abstract}

\maketitle

\section{Introduction}

Black-hole perturbations translate the geometry of a compact object into a set of characteristic wave equations. Their quasinormal modes connect exact background solutions with observable ringdown signals and, for this reason, have become one of the main tools of black-hole spectroscopy \cite{Kokkotas:1999bd,Berti:2009kk,Konoplya:2011qq,Bolokhov:2025uxz}. Different spin sectors probe the same spacetime in complementary ways: scalar fields are technically simple and geometrically transparent, electromagnetic perturbations isolate the gauge-field response, and Dirac fields offer a fermionic channel whose effective potential is structurally distinct from the bosonic ones. For massless fields in asymptotically flat spacetimes, all of these sectors share especially simple boundary conditions because the effective potentials vanish at both asymptotic ends.

Regular black holes provide a particularly appealing arena for such comparisons. Replacing the central singularity by a smooth core modifies the near-center geometry while keeping the exterior black-hole character, and this change can leave a measurable imprint on wave propagation, grey-body factors, and quasinormal ringing. A broad literature has already explored scalar, electromagnetic, gravitational, and Dirac perturbations of regular black holes in several models \cite{Flachi:2012nv,Yang:2021cvh,Konoplya:2022hll,Zhang:2024nny,Cai:2021ele,Li:2014fka,Konoplya:2023ahd,Guo:2024jhg,Jusufi:2020odz,Huang:2023aet,Lopez:2022uie,Fernando:2012yw,MahdavianYekta:2019pol,Gingrich:2024tuf,Konoplya:2023aph,Al-Badawi:2023lke,Jawad:2020hju,Konoplya:2026gim,DuttaRoy:2022ytr,Al-Badawi:2023lke,Pedraza:2021hzw,Lin:2013ofa,Mukohyama:2023xyf,Konoplya:2023ppx,Held:2019xde,Konoplya:2025ect,Konoplya:2025hgp,Saka:2025xxl,Bolokhov:2023ruj,Skvortsova:2026unq}. In that context, comparing several spin sectors within one and the same exact geometry is a natural way to identify which spectral features are universal and which are field dependent.

It has also been argued that strong-field observables such as quasinormal modes can depend predominantly on only a few effective parameters of the metric even within a very general black-hole parametrization; from that perspective, tracking the role of the single regularity scale $a$ in the present exact solution is especially well motivated \cite{Konoplya:2020pbh}.

Recently, Parvez and Shankaranarayanan obtained an exact asymptotically flat, nonsingular black-hole solution supported by a phantom Dirac--Born--Infeld (DBI) scalar \cite{Parvez:2025dbi}. The construction is attractive because the regular metric is not introduced ad hoc: it follows from a definite matter action, possesses scalar hair, and interpolates between black-hole, extremal-remnant, and horizonless regular configurations depending on the parameters \cite{Parvez:2025dbi}. The association between the DBI field and primordial dark matter arises from the fact that this field supports regular black-hole solutions in which the central singularity is replaced by a finite core due to the nonlinear kinetic structure. In such configurations, Hawking evaporation does not proceed to complete disappearance but instead terminates at a stable or long-lived remnant. As a result, primordial black holes formed in the early Universe could survive until the present epoch in the form of these remnants. Consequently, such DBI-supported black-hole relics provide a viable candidate for dark matter, linking modifications of the near-horizon and interior geometry to cosmological implications.  This makes the background a useful laboratory for testing how a nontrivial matter sector reshapes bosonic and fermionic wave dynamics. After the first version of this paper appeared on arXiv, a study of quasinormal modes of a massive scalar field in the DBI-supported black hole was presented in \cite{Skvortsova:2026jtx}, while the corresponding grey-body factors were analyzed in \cite{Bolokhov:2026kqu}.

In the present work we consider massless scalar, electromagnetic, and Dirac test fields in the DBI-supported regular black-hole geometry and analyze their quasinormal response within one common framework. After deriving the corresponding master equations and effective potentials, we compute representative quasinormal frequencies by means of the Pad\'e-improved WKB method and verify the scalar fundamental mode in the time domain. The results show that the regularity parameter systematically lowers and broadens the potential barriers, shifts the quasinormal frequencies away from the Schwarzschild limit, and modifies the damping/oscillation balance only mildly. At the same time, the spectral shifts are consistently larger than the estimated numerical uncertainty, so the spin-dependent imprint of the DBI regularity scale can be regarded as physically meaningful.

The paper is organized as follows. In Sec.~II we review the DBI-supported regular black-hole geometry. In Sec.~III we derive the master equations and effective potentials for the massless scalar, electromagnetic, and Dirac perturbations. In Sec.~IV we discuss the qualitative behavior of the corresponding potential barriers. In Sec.~V we outline the characteristic time-domain and Pad\'e-improved WKB methods used in the analysis. In Sec.~VI we present and interpret the quasinormal-mode results, emphasizing both their dependence on the regularity parameter and the accuracy of the numerical estimates. Finally, in Sec.~VII we summarize the main conclusions and outline possible extensions of the present work.

\section{Background geometry}\label{sec:geometry}

Following Ref.~\cite{Parvez:2025dbi}, we consider Einstein gravity coupled minimally to a non-linear Dirac--Born--Infeld scalar field. A convenient form of the action is
\begin{equation}\label{eq:action}
S=\int d^4x\sqrt{-g}\left[\frac{R}{16\pi G}-\epsilon\Lambda^4\left(\sqrt{1+\frac{\nabla_\mu\phi\nabla^\mu\phi}{\Lambda^4}}-1\right)\right],
\end{equation}
where $\Lambda$ sets the DBI scale and $\epsilon=+1$ or $-1$ distinguishes the canonical and phantom branches. The regular black-hole solution relevant for the present study is supported by the phantom branch \cite{Parvez:2025dbi}.

The static, spherically symmetric line element can be written as
\begin{equation}\label{eq:metric}
ds^2=-f(r)dt^2+\frac{dr^2}{f(r)}+R^2(r)(d\theta^2+\sin^2\theta\,d\phi^2),
\end{equation}
with
\begin{equation}\label{eq:metricfunctions}
\begin{aligned}
f(r)&=1+\frac{3M}{a}\left(\frac{r}{a}-\frac{a^2+r^2}{a^2}\arctan\frac{a}{r}\right),\\
R^2(r)&=r^2+a^2.
\end{aligned}
\end{equation}
Here $M$ is the ADM mass and $a$ sets the regularity scale. The areal radius never shrinks below $a$, so the center is replaced by a minimal two-sphere instead of a curvature singularity. At large distances one has $R(r)\sim r$ and
\begin{equation}
f(r)=1-\frac{2M}{r}+\Order{r^{-3}},
\end{equation}
which shows that the geometry is asymptotically flat. In the figures below we measure all dimensional quantities in units of the mass and set $M=1$.

The tortoise coordinate is defined in the standard way,
\begin{equation}\label{eq:tortoise}
dr_*\equiv\frac{dr}{f(r)}.
\end{equation}
For the black-hole branch, $r_*\to -\infty$ at the event horizon and $r_*\to +\infty$ at spatial infinity. This coordinate is the natural one for rewriting all perturbation equations in Schr\"odinger-like form.

\section{Massless test fields and effective potentials}\label{sec:fields}

For each spin sector considered here, the field equations reduce to a one-dimensional master equation,
\begin{equation}\label{eq:mastergeneral}
\frac{d^2\Psi}{dr_*^2}+\left(\omega^2-V(r)\right)\Psi=0,
\end{equation}
with a field-dependent effective potential $V(r)$. The different potentials reflect how the spin of the perturbing field couples to the same background geometry.

\subsection{Massless scalar field}

The massless scalar field obeys the Klein--Gordon equation
\begin{equation}\label{eq:kg}
\Box\Phi=0.
\end{equation}
Using the decomposition
\begin{equation}
\Phi(t,r,\theta,\phi)=e^{-\imo\omega t}Y_{\ell m}(\theta,\phi)\frac{\Psi_s(r)}{R(r)},
\end{equation}
one obtains Eq.~(\ref{eq:mastergeneral}) with the scalar potential
\begin{equation}\label{eq:Vscalarcompact}
V_s(r)=f(r)\left[\frac{\ell(\ell+1)}{R^2(r)}+\frac{1}{R(r)}\frac{d^2R(r)}{dr_*^2}\right],
\end{equation}
where $\ell=0,1,2,\ldots$. Since
\begin{equation}
f'(r)=\frac{6M}{a^3}\left(a-r\arctan\frac{a}{r}\right),
\end{equation}
and
\begin{equation}
\frac{1}{R}\frac{d^2R}{dr_*^2}=\frac{r f f'}{r^2+a^2}+\frac{a^2 f^2}{(r^2+a^2)^2},
\end{equation}
the scalar potential can also be written explicitly as
\begin{equation}\label{eq:Vscalarexplicit}
V_s(r)=f\left[\frac{\ell(\ell+1)}{r^2+a^2}+\frac{r f'}{r^2+a^2}+\frac{a^2f}{(r^2+a^2)^2}\right].
\end{equation}
For $a\to 0$ this expression reduces to the familiar Schwarzschild scalar potential.

Notice that the background black-hole solution is sourced by a non-minimally coupled scalar field. Nevertheless, the simplest model for perturbations considers a minimally coupled scalar field propagating in this background. The presence of two scalar sectors is quite common when studying field perturbations in black-hole spacetimes arising in theories with scalar degrees of freedom \cite{Fernando:2003wc,Konoplya:2002ky,Konoplya:2019hml,Bolokhov:2026dzn,Kokkotas:2015uma,Dubinsky:2024hmn}.

\subsection{Electromagnetic field}

The electromagnetic perturbation is governed by Maxwell's equations,
\begin{equation}\label{eq:maxwell}
\nabla_\mu F^{\mu\nu}=0,\qquad F_{\mu\nu}=\partial_\mu A_\nu-\partial_\nu A_\mu.
\end{equation}
After decomposing the vector potential into vector spherical harmonics, both parity sectors reduce to the same master equation, Eq.~(\ref{eq:mastergeneral}), with the effective potential
\begin{equation}\label{eq:Vem}
V_{em}(r)=f(r)\frac{\ell(\ell+1)}{R^2(r)}=f(r)\frac{\ell(\ell+1)}{r^2+a^2},
\end{equation}
valid for $\ell=1,2,3,\ldots$. In comparison with the scalar case, the electromagnetic barrier is purely centrifugal and does not contain an additional curvature term.

\subsection{Massless Dirac field}

For the massless Dirac field we start from
\begin{equation}\label{eq:dirac}
\gamma^\mu\nabla_\mu\Upsilon=0.
\end{equation}
Separating the angular dependence in spinor spherical harmonics leads to two radial functions whose first-order system can be written as
\begin{equation}\label{eq:diracfirstorder}
\left(\frac{d}{dr_*}-W\right)F=-\imo\omega G,\qquad
\left(\frac{d}{dr_*}+W\right)G=-\imo\omega F,
\end{equation}
with the superpotential
\begin{equation}\label{eq:superpotential}
W(r)=\frac{|\kappa|\sqrt{f(r)}}{R(r)}=\frac{|\kappa|\sqrt{f(r)}}{\sqrt{r^2+a^2}},
\end{equation}
where $\kappa=\pm1,\pm2,\ldots$ is the usual separation constant. Introducing $Z_{\pm}=F\pm G$, one obtains two supersymmetric partner equations,
\begin{equation}\label{eq:diracmaster}
\frac{d^2Z_{\pm}}{dr_*^2}+\left(\omega^2-V_{\pm}(r)\right)Z_{\pm}=0,
\end{equation}
with
\begin{equation}\label{eq:Vdiraccompact}
V_{\pm}(r)=W^2\pm\frac{dW}{dr_*}.
\end{equation}
An explicit form is
\begin{equation}\label{eq:Vdiracexplicit}
V_{\pm}(r)=\frac{\kappa^2f}{r^2+a^2}\pm |\kappa|\sqrt{f}\left[\frac{f'}{2\sqrt{r^2+a^2}}-\frac{fr}{(r^2+a^2)^{3/2}}\right].
\end{equation}
The two partner potentials are isospectral, so throughout the discussion of the plots we display only $V_+$.

\section{Representative potential profiles}\label{sec:potentials}

All three perturbation sectors share the same qualitative asymptotics in the black-hole regime: the effective potential vanishes at the horizon and also tends to zero at spatial infinity. Consequently, the problem has the standard quasinormal-mode boundary conditions of purely ingoing waves at $r_*\to -\infty$ and purely outgoing waves at $r_*\to +\infty$. The difference between the fields is encoded in the shape and height of the intermediate barrier.

\begin{figure}
\centering
\includegraphics[width=\linewidth]{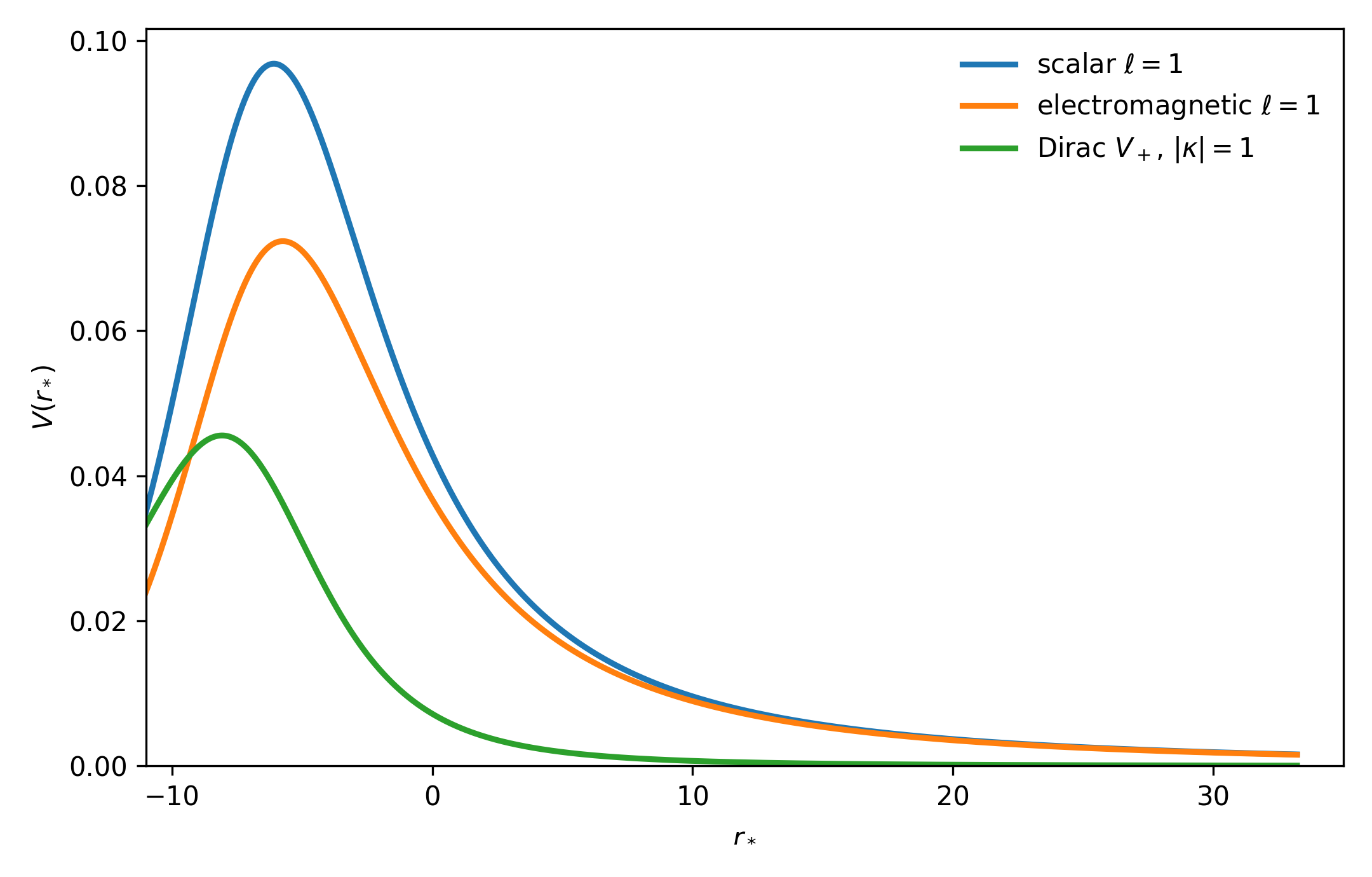}
\caption{Effective potentials in the DBI-supported regular black-hole background for $M=1$ and $a=0.6$: massless scalar field with $\ell=1$, electromagnetic field with $\ell=1$, and the Dirac partner potential $V_+$ with $|\kappa|=1$. The scalar barrier is highest, while the Dirac barrier peaks closer to the horizon.}
\label{fig:compare}
\end{figure}

Figure~\ref{fig:compare} compares representative barriers for the scalar, electromagnetic, and Dirac sectors at $a=0.6$ and $M=1$. For the same low multipole, the scalar barrier lies above the electromagnetic one because of the additional geometric term in Eq.~(\ref{eq:Vscalarcompact}). The Dirac barrier shown through $V_+$ is lower still and peaks closer to the horizon. Thus, even before calculating the spectrum, one already sees that the spin of the test field should affect both the oscillation frequency and the damping rate.

The influence of the regularity scale is illustrated in Fig.~\ref{fig:areg}. In all three sectors, increasing $a$ lowers the maximum of the barrier and makes the profile slightly broader. This tendency is mild but systematic, and it is precisely the type of deformation that is expected to shift the quasinormal frequencies relative to the Schwarzschild limit. In particular, the DBI regularity scale modifies not only the height of the bosonic barriers but also the fermionic superpartner potentials, which makes a spin-by-spin comparison especially interesting.

\begin{figure}
\centering
\includegraphics[width=\linewidth]{massless_fields_regularities.png}
\caption{Dependence of the effective potentials on the regularity parameter $a$ for $M=1$. From top to bottom: scalar potential with $\ell=1$, electromagnetic potential with $\ell=1$, and Dirac partner potential $V_+$ with $|\kappa|=1$. In each sector, increasing $a$ lowers and broadens the barrier.}
\label{fig:areg}
\end{figure}

The single-barrier structure seen in these profiles is also important methodologically. It indicates that both high-order WKB techniques and characteristic time-domain integration can be applied in a straightforward way once the spectral calculations are carried out.

\section{Methods}\label{sec:methods}

To determine the quasinormal frequencies in the next stage of the project, we will use two complementary approaches. We begin with direct characteristic evolution in the time domain, which provides the full waveform and an independent check of the dominant ringing. We then turn to the high-order WKB expansion improved by Pad\'e resummation, which offers an efficient semi-analytic estimate of the spectrum.

\noindent\textbf{Time-domain integration.}
We begin with characteristic integration in null coordinates,
\begin{equation}
u=t-r_*,\qquad v=t+r_*.
\end{equation}
In these coordinates the master equations are evolved with the Gundlach--Price--Pullin scheme \cite{Gundlach:1993tp},
\begin{eqnarray}\label{eq:gpp}
\Psi(N)&=&\Psi(W)+\Psi(E)-\Psi(S)\nonumber\\
&&-\frac{\Delta^2}{8}V(S)\left[\Psi(W)+\Psi(E)\right]+\Order{\Delta^4},
\end{eqnarray}
where $N=(u+\Delta,v+\Delta)$, $W=(u+\Delta,v)$, $E=(u,v+\Delta)$, and $S=(u,v)$. This evolution reconstructs the prompt signal, the exponentially damped ringdown, and the late-time tail within one numerical framework. The dominant frequencies may then be extracted from the ringdown stage, for example by a Prony fit to a finite sum of damped exponentials \cite{Berti:2009kk,Konoplya:2011qq}. Time-domain integration method is frequently used for finding dominant quasinormal modes and testing stability of perturbations \cite{Bolokhov:2023bwm,Malik:2023bxc,Skvortsova:2024atk,Arbelaez:2025gwj,Konoplya:2013sba,Bolokhov:2023dxq,Dubinsky:2024nzo,Varghese:2011ku,Skvortsova:2023zmj,Bolokhov:2026eqf,Abdalla:2012si,Konoplya:2024hfg,Arbelaez:2026eaz,Skvortsova:2024wly,Malik:2024nhy,Dubinsky:2025fwv,Konoplya:2023fmh,Bolokhov:2024ixe,Lutfuoglu:2025hjy}. Here we use it to verify the results obtained via the WKB method. The latter, being significantly less computationally demanding, serves as our primary approach.

\medskip

\noindent\textbf{WKB approximation.}
As a complementary semi-analytic method, we use the WKB approach. For massless fields, the quasinormal boundary conditions take the form
\begin{equation}\label{eq:bc}
\Psi(r_*\to-\infty)\propto e^{-\imo\omega r_*},\qquad
\Psi(r_*\to+\infty)\propto e^{+\imo\omega r_*}.
\end{equation}
When the effective potential has a single smooth maximum, these asymptotic solutions can be matched to a local expansion near the peak. For spherically symmetric spacetimes, the WKB result may be organized as \cite{Iyer:1986np,Konoplya:2003ii,Matyjasek:2017psv,Matyjasek:2026yiu}
\begin{equation}\label{eq:wkb}
\begin{aligned}
\omega^2 ={}& V_0 + \sum_{j=1}^{\infty} A_{2j}(\K^2) \\
& {} - \imo\K\sqrt{-2V_2}\left[1+\sum_{j=1}^{\infty} A_{2j+1}(\K^2)\right].
\end{aligned}
\end{equation}
where
\begin{equation}
\K=n+\frac{1}{2},\qquad n=0,1,2,\ldots.
\end{equation}
Here $V_0$ is the value of the potential at its maximum and $V_2$ is the second derivative with respect to $r_*$ at that point. For the Dirac field the same procedure can be applied to either of the isospectral partner potentials. In practical calculations we will use Pad\'e-improved versions of the high-order WKB series, since these often provide a more stable estimate for low multipoles and the fundamental overtone. The WKB method at various orders have been applied in numerous works for black holes and other compact objects \cite{Konoplya:2010vz,Malik:2025erb,Guo:2020caw,Kokkotas:2010zd,Konoplya:2009hv,Skvortsova:2024msa,Abdalla:2005hu,Bolokhov:2025lnt,Konoplya:2006ar,Pathrikar:2025gzu,Ishihara:2008re,Bolokhov:2025aqy,Konoplya:2018ala,Fernando:2016ftj,Momennia:2018hsm,Eniceicu:2019npi,Konoplya:2019ppy,Wongjun:2019ydo,Malik:2026lfj,Breton:2017hwe,Karmakar:2023cwg,Konoplya:2001ji,Konoplya:2005sy,Bolokhov:2025egl,Lutfuoglu:2025ljm} and showed very good concordance with more accurate method whenever $\ell \geq n$.

Together, the time-domain and WKB approaches provide a robust strategy for the spectral analysis developed below.

\section{Quasinormal modes}

In this section we list the fundamental frequencies ($n=0$) obtained from the Pad\'e-improved WKB approach for representative values of the regularity parameter. As an internal consistency check, we compare the 16th-order WKB results with Pad\'e approximant $\tilde{m}=8$ against the 14th-order results with $\tilde{m}=7$. For scalar perturbations with $\ell=1$ and $\ell=2$ at $a=0.2$, the effective potential has no barrier maximum, and in those cases the WKB method is not applicable.

To complement the tabulated frequencies, it is useful to track the quality factor $Q=\re{\omega}/(2|\im{\omega}|)$ computed from the WKB16 values. This quantity measures the balance between oscillation and damping: larger $Q$ corresponds to a longer-lived and more oscillatory ringdown.

\begin{table}[t]
\caption{Fundamental quasinormal frequencies ($n=0$) of the massless scalar field in the asymptotically flat regular black-hole background for $M=1$. Each panel compares the 16th-order WKB result improved by the Pad\'e approximant $\tilde{m}=8$ with the 14th-order result improved by $\tilde{m}=7$; the last column gives their relative difference in percent. Entries marked ``no peak found'' indicate that the effective potential does not possess a barrier maximum, so the WKB method cannot be applied. Here and in the following tables  "0\%" indicate agreement only within the displayed six digits.}
\label{tab:scalar-qnms}
\centering
\small
\begin{tabular}{c c c c}
\hline
\multicolumn{4}{c}{(a) $\ell=0$}\\
\hline
$a$ & WKB16 ($\tilde{m}=8$) & WKB14 ($\tilde{m}=7$) & difference \\
\hline
$0.2$ & $0.110223-0.104792 i$ & $0.110146-0.104450 i$ & $0.230\%$\\
$0.3$ & $0.109911-0.104592 i$ & $0.109837-0.104249 i$ & $0.231\%$\\
$0.4$ & $0.109478-0.104313 i$ & $0.109409-0.103967 i$ & $0.233\%$\\
$0.5$ & $0.108925-0.103959 i$ & $0.108865-0.103603 i$ & $0.239\%$\\
$0.6$ & $0.108264-0.103538 i$ & $0.108206-0.103167 i$ & $0.251\%$\\
$0.7$ & $0.107523-0.103067 i$ & $0.107425-0.102705 i$ & $0.252\%$\\
$0.8$ & $0.106658-0.102505 i$ & $0.106547-0.102221 i$ & $0.206\%$\\
$0.9$ & $0.105599-0.101808 i$ & $0.105594-0.101617 i$ & $0.130\%$\\
$1.$ & $0.104535-0.101138 i$ & $0.104552-0.100903 i$ & $0.163\%$\\
\hline
\end{tabular}

\medskip

\begin{tabular}{c c c c}
\hline
\multicolumn{4}{c}{(b) $\ell=1$}\\
\hline
$a$ & WKB16 ($\tilde{m}=8$) & WKB14 ($\tilde{m}=7$) & difference \\
\hline
%$0.2$ & \multicolumn{3}{c}{no peak found}\\
$0.4$ & $0.291260-0.097149 i$ & $0.291260-0.097149 i$ & $0\%$\\
$0.6$ & $0.289214-0.096526 i$ & $0.289214-0.096526 i$ & $0\%$\\
$0.8$ & $0.286438-0.095681 i$ & $0.286438-0.095681 i$ & $<10^{-4}\%$\\
$1.$ & $0.283009-0.094640 i$ & $0.283009-0.094640 i$ & $0\%$\\
$1.2$ & $0.279019-0.093429 i$ & $0.279019-0.093429 i$ & $0\%$\\
$1.4$ & $0.274561-0.092079 i$ & $0.274561-0.092079 i$ & $<10^{-4}\%$\\
$1.6$ & $0.269730-0.090618 i$ & $0.269730-0.090618 i$ & $0\%$\\
$1.8$ & $0.264614-0.089073 i$ & $0.264614-0.089073 i$ & $0\%$\\
$2.$ & $0.259297-0.087470 i$ & $0.259297-0.087469 i$ & $0.0003\%$\\
\hline
\end{tabular}

\medskip

\begin{tabular}{c c c c}
\hline
\multicolumn{4}{c}{(c) $\ell=2$}\\
\hline
$a$ & WKB16 ($\tilde{m}=8$) & WKB14 ($\tilde{m}=7$) & difference \\
\hline
%$0.2$ & \multicolumn{3}{c}{no peak found}\\
$0.4$ & $0.481021-0.096255 i$ & $0.481021-0.096255 i$ & $0\%$\\
$0.6$ & $0.477816-0.095639 i$ & $0.477816-0.095639 i$ & $0\%$\\
$0.8$ & $0.473462-0.094802 i$ & $0.473462-0.094802 i$ & $0\%$\\
$1.$ & $0.468076-0.093767 i$ & $0.468076-0.093767 i$ & $0\%$\\
$1.2$ & $0.461793-0.092560 i$ & $0.461793-0.092560 i$ & $0\%$\\
$1.4$ & $0.454758-0.091209 i$ & $0.454758-0.091209 i$ & $0\%$\\
$1.6$ & $0.447113-0.089741 i$ & $0.447113-0.089741 i$ & $0\%$\\
$1.8$ & $0.438997-0.088183 i$ & $0.438997-0.088183 i$ & $0\%$\\
$2.$ & $0.430536-0.086557 i$ & $0.430536-0.086557 i$ & $0\%$\\
\hline
\end{tabular}
\end{table}

\begin{table}[t]
\caption{Fundamental quasinormal frequencies ($n=0$) of the electromagnetic field in the asymptotically flat regular black-hole background for $M=1$. The two panels correspond to $\ell=1$ and $\ell=2$. In each case we compare the 16th-order and 14th-order WKB results improved by Pad\'e approximants with $\tilde{m}=8$ and $\tilde{m}=7$, respectively; the last column gives their relative difference in percent.}
\label{tab:maxwell-qnms}
\centering
\small
\begin{tabular}{c c c c}
\hline
\multicolumn{4}{c}{(a) $\ell=1$}\\
\hline
$a$ & WKB16 ($\tilde{m}=8$) & WKB14 ($\tilde{m}=7$) & difference \\
\hline
$0.2$ & $0.247934-0.092359 i$ & $0.247934-0.092359 i$ & $0\%$\\
$0.4$ & $0.246954-0.091978 i$ & $0.246954-0.091978 i$ & $0\%$\\
$0.6$ & $0.245354-0.091356 i$ & $0.245354-0.091356 i$ & $0\%$\\
$0.8$ & $0.243177-0.090509 i$ & $0.243177-0.090509 i$ & $0\%$\\
$1.$ & $0.240480-0.089460 i$ & $0.240480-0.089461 i$ & $0.0001\%$\\
$1.2$ & $0.237329-0.088235 i$ & $0.237329-0.088236 i$ & $0.0001\%$\\
$1.4$ & $0.233794-0.086862 i$ & $0.233794-0.086862 i$ & $0.0001\%$\\
$1.6$ & $0.229946-0.085367 i$ & $0.229946-0.085367 i$ & $0\%$\\
$1.8$ & $0.225851-0.083777 i$ & $0.225850-0.083778 i$ & $<10^{-4}\%$\\
$2.$ & $0.221572-0.082118 i$ & $0.221572-0.082118 i$ & $0.0001\%$\\
\hline
\end{tabular}

\medskip

\begin{tabular}{c c c c}
\hline
\multicolumn{4}{c}{(b) $\ell=2$}\\
\hline
$a$ & WKB16 ($\tilde{m}=8$) & WKB14 ($\tilde{m}=7$) & difference \\
\hline
$0.2$ & $0.456985-0.094875 i$ & $0.456985-0.094875 i$ & $0\%$\\
$0.4$ & $0.455172-0.094491 i$ & $0.455172-0.094491 i$ & $0\%$\\
$0.6$ & $0.452208-0.093863 i$ & $0.452208-0.093863 i$ & $0\%$\\
$0.8$ & $0.448177-0.093010 i$ & $0.448177-0.093010 i$ & $0\%$\\
$1.$ & $0.443185-0.091953 i$ & $0.443185-0.091953 i$ & $0\%$\\
$1.2$ & $0.437355-0.090719 i$ & $0.437355-0.090719 i$ & $0\%$\\
$1.4$ & $0.430815-0.089336 i$ & $0.430815-0.089336 i$ & $0\%$\\
$1.6$ & $0.423697-0.087830 i$ & $0.423697-0.087830 i$ & $0\%$\\
$1.8$ & $0.416126-0.086230 i$ & $0.416126-0.086230 i$ & $0\%$\\
$2.$ & $0.408219-0.084559 i$ & $0.408219-0.084559 i$ & $0\%$\\
\hline
\end{tabular}
\end{table}

\begin{table}[t]
\caption{Fundamental quasinormal frequencies ($n=0$) of the massless Dirac field in the asymptotically flat regular black-hole background for $M=1$. The two panels correspond to the angular separation constants $|\kappa|=1$ and $|\kappa|=2$. In each case we compare the 16th-order and 14th-order WKB results improved by Pad\'e approximants with $\tilde{m}=8$ and $\tilde{m}=7$, respectively; the last column gives their relative difference in percent.}
\label{tab:dirac-qnms}
\centering
\small
\begin{tabular}{c c c c}
\hline
\multicolumn{4}{c}{(a) $|\kappa|=1$}\\
\hline
$a$ & WKB16 ($\tilde{m}=8$) & WKB14 ($\tilde{m}=7$) & difference \\
\hline
$0.2$ & $0.182519-0.097163 i$ & $0.182603-0.096972 i$ & $0.101\%$\\
$0.4$ & $0.181816-0.096789 i$ & $0.181867-0.096573 i$ & $0.108\%$\\
$0.6$ & $0.180686-0.096164 i$ & $0.180664-0.095919 i$ & $0.120\%$\\
$0.8$ & $0.179151-0.095255 i$ & $0.179028-0.095027 i$ & $0.127\%$\\
$1.$ & $0.177181-0.094080 i$ & $0.177002-0.093919 i$ & $0.120\%$\\
$1.2$ & $0.174805-0.092730 i$ & $0.174632-0.092619 i$ & $0.104\%$\\
$1.4$ & $0.172111-0.091248 i$ & $0.171969-0.091159 i$ & $0.0860\%$\\
$1.6$ & $0.169172-0.089648 i$ & $0.169066-0.089569 i$ & $0.0688\%$\\
$1.8$ & $0.166052-0.087948 i$ & $0.165977-0.087879 i$ & $0.0544\%$\\
$2.$ & $0.162804-0.086176 i$ & $0.162750-0.086116 i$ & $0.0438\%$\\
\hline
\end{tabular}

\medskip

\begin{tabular}{c c c c}
\hline
\multicolumn{4}{c}{(b) $|\kappa|=2$}\\
\hline
$a$ & WKB16 ($\tilde{m}=8$) & WKB14 ($\tilde{m}=7$) & difference \\
\hline
$0.2$ & $0.379527-0.096275 i$ & $0.379527-0.096275 i$ & $0\%$\\
$0.4$ & $0.378014-0.095887 i$ & $0.378014-0.095887 i$ & $0\%$\\
$0.6$ & $0.375540-0.095255 i$ & $0.375540-0.095255 i$ & $0\%$\\
$0.8$ & $0.372176-0.094394 i$ & $0.372176-0.094394 i$ & $0\%$\\
$1.$ & $0.368009-0.093328 i$ & $0.368010-0.093328 i$ & $0\%$\\
$1.2$ & $0.363143-0.092084 i$ & $0.363143-0.092084 i$ & $0\%$\\
$1.4$ & $0.357686-0.090688 i$ & $0.357686-0.090688 i$ & $0\%$\\
$1.6$ & $0.351746-0.089170 i$ & $0.351746-0.089170 i$ & $0\%$\\
$1.8$ & $0.345429-0.087556 i$ & $0.345429-0.087556 i$ & $0\%$\\
$2.$ & $0.338832-0.085870 i$ & $0.338832-0.085870 i$ & $0\%$\\
\hline
\end{tabular}
\end{table}

\begin{table}[t]
\caption{Higher-overtone quasinormal frequencies of the electromagnetic field in the asymptotically flat regular black-hole background for $M=1$ and $\ell=2$. The three panels correspond to the overtones $n=1$, $n=2$, and $n=3$. In each case we compare the 16th-order WKB result improved by the Pad\'e approximant $\tilde{m}=8$ with the 14th-order result improved by $\tilde{m}=7$; the last column gives their relative difference in percent.}
\label{tab:maxwell-higher-qnms}
\centering
\small
\begin{tabular}{c c c c}
\hline
\multicolumn{4}{c}{(a) $n=1$}\\
\hline
$a$ & WKB16 ($\tilde{m}=8$) & WKB14 ($\tilde{m}=7$) & difference \\
\hline
$0.2$ & $0.435963-0.290314 i$ & $0.435963-0.290314 i$ & $0\%$\\
$0.4$ & $0.434242-0.289139 i$ & $0.434242-0.289139 i$ & $0\%$\\
$0.6$ & $0.431428-0.287217 i$ & $0.431428-0.287217 i$ & $0\%$\\
$0.8$ & $0.427598-0.284602 i$ & $0.427598-0.284602 i$ & $0\%$\\
$1.$ & $0.422851-0.281364 i$ & $0.422851-0.281364 i$ & $0\%$\\
$1.2$ & $0.417301-0.277582 i$ & $0.417301-0.277582 i$ & $0\%$\\
$1.4$ & $0.411071-0.273338 i$ & $0.411071-0.273338 i$ & $0\%$\\
$1.6$ & $0.404282-0.268719 i$ & $0.404282-0.268719 i$ & $0\%$\\
$1.8$ & $0.397053-0.263805 i$ & $0.397053-0.263805 i$ & $0\%$\\
$2.$ & $0.389494-0.258673 i$ & $0.389494-0.258673 i$ & $0\%$\\
\hline
\end{tabular}

\medskip

\begin{tabular}{c c c c}
\hline
\multicolumn{4}{c}{(b) $n=2$}\\
\hline
$a$ & WKB16 ($\tilde{m}=8$) & WKB14 ($\tilde{m}=7$) & difference \\
\hline
$0.2$ & $0.400661-0.500903 i$ & $0.400659-0.500902 i$ & $0.00021\%$\\
$0.4$ & $0.399096-0.498871 i$ & $0.399096-0.498871 i$ & $0.00010\%$\\
$0.6$ & $0.396537-0.495550 i$ & $0.396536-0.495550 i$ & $0.00013\%$\\
$0.8$ & $0.393048-0.491030 i$ & $0.393047-0.491029 i$ & $0.00030\%$\\
$1.$ & $0.388715-0.485428 i$ & $0.388714-0.485426 i$ & $0.00039\%$\\
$1.2$ & $0.383635-0.478876 i$ & $0.383634-0.478875 i$ & $0.0001\%$\\
$1.4$ & $0.377916-0.471522 i$ & $0.377915-0.471522 i$ & $0.00023\%$\\
$1.6$ & $0.371667-0.463508 i$ & $0.371662-0.463509 i$ & $0.00088\%$\\
$1.8$ & $0.364993-0.454977 i$ & $0.364985-0.454977 i$ & $0.00137\%$\\
$2.$ & $0.357993-0.446060 i$ & $0.357984-0.446058 i$ & $0.00165\%$\\
\hline
\end{tabular}

\medskip

\begin{tabular}{c c c c}
\hline
\multicolumn{4}{c}{(c) $n=3$}\\
\hline
$a$ & WKB16 ($\tilde{m}=8$) & WKB14 ($\tilde{m}=7$) & difference \\
\hline
$0.2$ & $0.362130-0.729199 i$ & $0.362123-0.729224 i$ & $0.00318\%$\\
$0.4$ & $0.360742-0.726235 i$ & $0.360707-0.726255 i$ & $0.00496\%$\\
$0.6$ & $0.358468-0.721388 i$ & $0.358431-0.721398 i$ & $0.00479\%$\\
$0.8$ & $0.355357-0.714786 i$ & $0.355319-0.714792 i$ & $0.00482\%$\\
$1.$ & $0.351473-0.706597 i$ & $0.351437-0.706599 i$ & $0.00447\%$\\
$1.2$ & $0.346887-0.697013 i$ & $0.346864-0.697016 i$ & $0.00295\%$\\
$1.4$ & $0.341688-0.686252 i$ & $0.341684-0.686254 i$ & $0.00059\%$\\
$1.6$ & $0.335986-0.674524 i$ & $0.335984-0.674525 i$ & $0.00035\%$\\
$1.8$ & $0.329866-0.662016 i$ & $0.329848-0.662035 i$ & $0.00350\%$\\
$2.$ & $0.323403-0.648926 i$ & $0.323354-0.648967 i$ & $0.00882\%$\\
\hline
\end{tabular}
\end{table}

The tabulated quasinormal spectra reveal a clear and nearly universal dependence on the regularity parameter $a$. Since the DBI geometry reduces to the Schwarzschild solution in the limit $a\to 0$, increasing $a$ measures the departure from the Schwarzschild case. For all fundamental modes displayed in Tabs.~\ref{tab:scalar-qnms}--\ref{tab:dirac-qnms}, both the oscillation frequency $\re{\omega}$ and the damping rate $|\im{\omega}|$ decrease monotonically as $a$ grows. In the complex-frequency plane this means that the modes drift toward the origin: the ringing becomes simultaneously less oscillatory and less strongly damped. The size of the effect is not marginal. For example, for the scalar mode with $\ell=1$ one has $\omega=0.291260-0.097149\,i$ at $a=0.4$ and $\omega=0.259297-0.087470\,i$ at $a=2$, which corresponds to shifts of about $11\%$ in $\re{\omega}$ and $10\%$ in $|\im{\omega}|$. The electromagnetic mode with $\ell=1$ and the Dirac mode with $|\kappa|=1$ exhibit a very similar trend, with changes of about $(10.6,11.1)\%$ and $(10.8,11.3)\%$ in $(\re{\omega},|\im{\omega}|)$, respectively. Therefore, the dominant influence of the DBI regularity scale is qualitatively the same in all three spin sectors, although the absolute position of the modes remains spin dependent.

\begin{figure}[t]
\centering
\includegraphics[width=\linewidth]{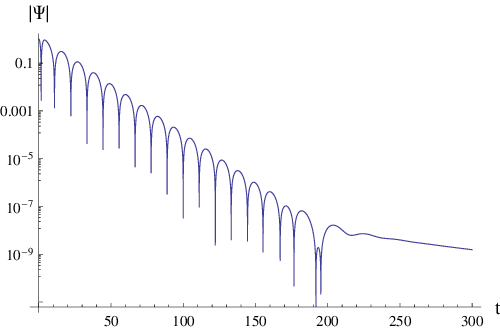}
\caption{Time-domain profile of the massless scalar perturbation for $\ell=1$, $M=1$, and $a=1$. A Prony fit to the ringdown gives $\omega=0.28301-0.0946395\,i$. For comparison, the WKB value from Tab.~\ref{tab:scalar-qnms} is $\omega=0.283009-0.094640\,i$, so the agreement is better than $6\times 10^{-4}\%$ in both the real and imaginary parts.}
\label{fig:td-scalar}
\end{figure}

The significance of this shift becomes especially clear once it is compared with the numerical uncertainty estimated from the difference between the WKB16 and WKB14 Pad\'e-improved results. For the fundamental modes with $\ell\geq 1$ or $|\kappa|\geq 2$, this difference is typically zero in the quoted digits and, when nonzero, remains at the level $10^{-4}\%$--$10^{-3}\%$. Even for the higher electromagnetic overtones in Tab.~\ref{tab:maxwell-higher-qnms}, the discrepancy does not exceed $8.82\times 10^{-3}\%$. The least favorable fundamental cases are the Dirac mode with $|\kappa|=1$, where the internal spread reaches about $0.13\%$, and the scalar monopole, where it is about $0.25\%$. Yet even in these cases the physical effect of changing $a$ is much larger: for the Dirac $|\kappa|=1$ mode the cumulative shift across the table is about $11\%$, while for the scalar monopole it is about $5.2\%$ in $\re{\omega}$ and $3.5\%$ in $|\im{\omega}|$. Thus, within the range where the WKB method is applicable, the deviation from the Schwarzschild spectrum is always at least one order of magnitude larger than the estimated numerical error, and for most modes it is larger by several orders of magnitude. In this sense we are on solid ground: the observed dependence on $a$ is not a numerical artifact but a genuine physical effect. This conclusion is strengthened by the time-domain check in Fig.~\ref{fig:td-scalar}, where for the scalar mode with $\ell=1$ at $a=1$ the Prony fit agrees with the WKB value to better than $6\times 10^{-4}\%$.

\begin{figure}[t]
\centering
\includegraphics[width=\linewidth]{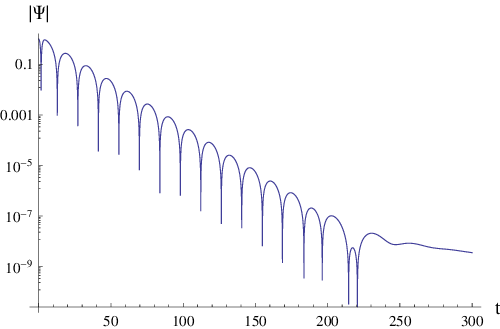}
\caption{Time-domain profile of the electromagnetic perturbation for $\ell=1$, $M=1$, and $a=2$. A Prony fit to the ringdown gives $\omega=0.221573-0.0821174\,i$. For comparison, the WKB value from Tab.~\ref{tab:maxwell-qnms} is $\omega=0.221572-0.082118\,i$, so the agreement is better than $7.3\times 10^{-4}\%$ in both the real and imaginary parts.}
\label{fig:td-em}
\end{figure}

As an independent time-domain verification of the scalar fundamental mode, Fig.~\ref{fig:td-scalar} shows the waveform for $\ell=1$, $M=1$, and $a=1$, for which the Prony fit is in excellent agreement with the WKB result listed in Tab.~\ref{tab:scalar-qnms}. An analogous electromagnetic profile is shown in Fig.~\ref{fig:td-em} for $\ell=1$, $M=1$, and $a=2$. In this case the Prony fit gives $\omega=0.221573-0.0821174\,i$, in excellent agreement with the WKB value $\omega=0.221572-0.082118\,i$ from Tab.~\ref{tab:maxwell-qnms}.

\begin{figure}[t]
\centering
\includegraphics[width=\linewidth]{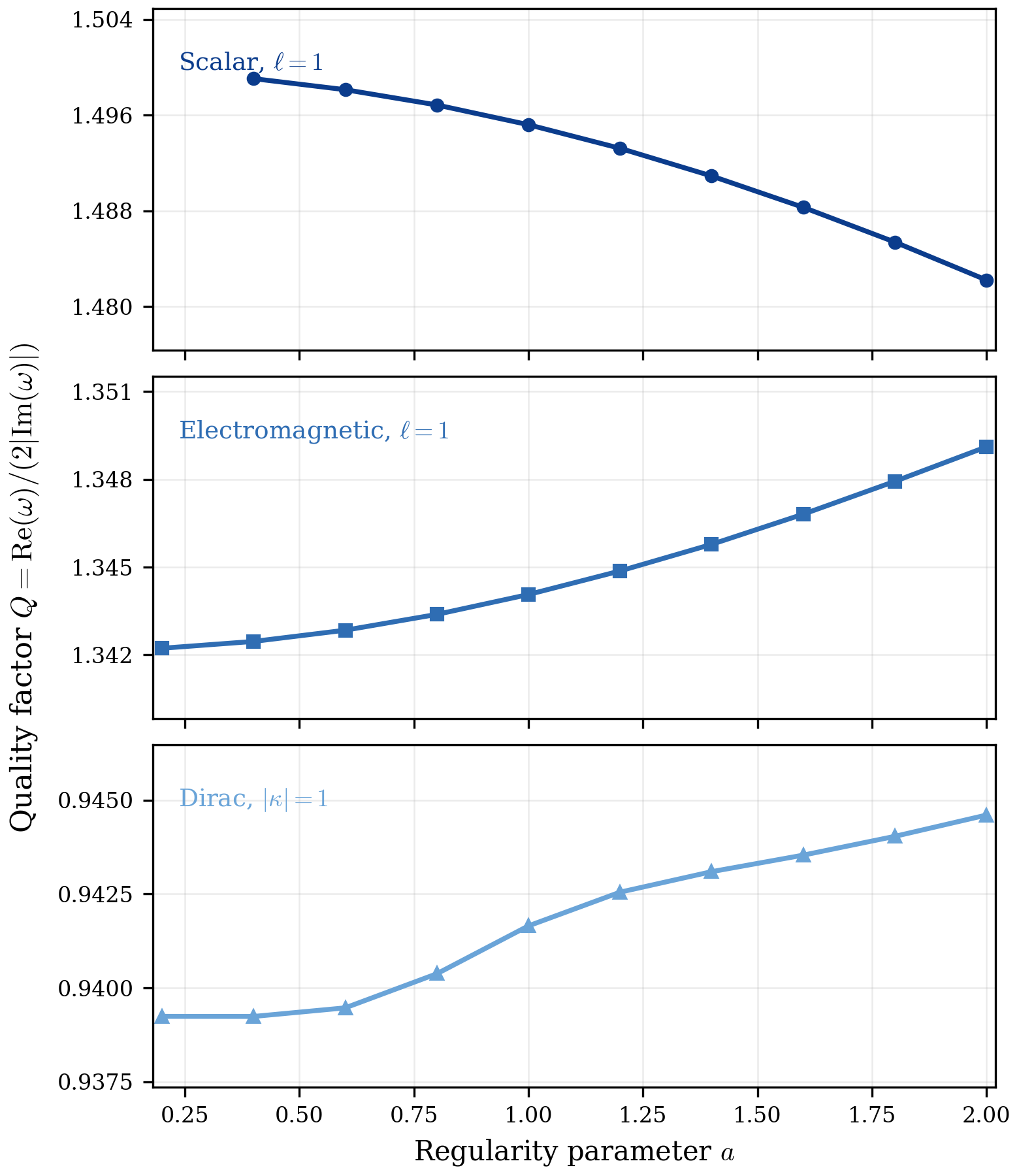}
\caption{Quality factor $Q=\re{\omega}/(2|\im{\omega}|)$ for representative fundamental modes extracted from the WKB16 values in Tabs.~\ref{tab:scalar-qnms}--\ref{tab:dirac-qnms}: scalar $\ell=1$, electromagnetic $\ell=1$, and Dirac $|\kappa|=1$. The three panels use separate vertical scales in order to make the weak dependence on $a$ clearly visible.}
\label{fig:qfactor}
\end{figure}

Figure~\ref{fig:qfactor} shows a weak but systematic dependence of the quality factor on the regularity scale. In the scalar sector $Q$ decreases with $a$ (for $\ell=1$, from $Q\simeq 1.499$ at $a=0.4$ to $Q\simeq 1.482$ at $a=2.0$), whereas the electromagnetic and Dirac cases increase slightly (from $Q\simeq 1.342$ to $1.349$ and from $Q\simeq 0.939$ to $0.945$, respectively). Thus the DBI regularity parameter changes the damping/oscillation balance only mildly, while larger angular numbers still correspond to larger $Q$ within each spin sector.

For the massless fields considered here, the quasinormal spectrum is not only useful for ringdown itself: it can also be employed to approximate the corresponding grey-body factors. The reason is that, for single-barrier effective potentials, the WKB scattering quantity $\K_{\ell}(\Omega)$ entering the transmission probability can be related directly to the dominant quasinormal frequencies. In the standard notation the partial grey-body factor is written as
\begin{equation}
\Gamma_{\ell}(\Omega)=|T_{\ell}(\Omega)|^2=\left(1+e^{2\pi i\K_{\ell}(\Omega)}\right)^{-1},
\end{equation}
with $\Omega$ denoting the real scattering frequency. According to the correspondence derived in Eqs.~(3.5)--(4.6) of Refs.~\cite{Konoplya:2024gbf,Konoplya:2024vuj}, the eikonal limit already allows one to reconstruct $\K_{\ell}(\Omega)$ from the fundamental mode $\omega_{\ell 0}$ alone, whereas beyond that limit a more accurate estimate is obtained by supplementing $\omega_{\ell 0}$ with the first overtone $\omega_{\ell 1}$. For the real scattering frequency ($\Omega)$, the correspondence can be written in the following way \cite{Konoplya:2024gbf,Konoplya:2024vuj,Bolokhov:2024otn}
\begin{equation}\label{transmission-eikonal}
\Gamma_{\ell}(\Omega) =
\left(1+e^{2\pi\dfrac{\Omega^2-\re{\omega_0}^2}{4\re{\omega_0}\im{\omega_0}}}\right)^{-1}
+ \Order{\ell^{-1}},
\end{equation} 
where the beyond eikonal corrections contain the first overtone.

Thus, once these lowest quasinormal frequencies are known, one can infer the transmission coefficients semi-analytically without solving an additional scattering problem.

This quasinormal-mode/grey-body-factor connection has already been explored in a variety of recent settings \cite{Lutfuoglu:2026gey,Lutfuoglu:2025mqa,Lutfuoglu:2025ldc,Malik:2024wvs,Lutfuoglu:2025eik,Konoplya:2010vz,Dubinsky:2024vbn,Lutfuoglu:2026uzy,Malik:2025qnr,Lutfuoglu:2025kqp,Bolokhov:2026kqu}, and the reported agreement is often good even for moderately small multipole numbers. At the same time, its applicability is tied to the same structural assumption that makes the WKB treatment reliable, namely the presence of a single smooth barrier. If the effective potential develops a double-well profile \cite{Konoplya:2025hgp}, or if higher-curvature corrections substantially deform the centrifugal sector and trigger catastrophic instabilities \cite{Konoplya:2017zwo,Takahashi:2010gz,Dotti:2004sh,Konoplya:2017ymp,Konoplya:2017lhs}, the correspondence is expected to deteriorate.

\section{Summary and outlook}\label{sec:summary}

We have investigated massless scalar, electromagnetic, and Dirac test-field perturbations of the asymptotically flat regular black-hole solution supported by a phantom Dirac--Born--Infeld scalar. For all three spin sectors the perturbation problem can be reduced to a Schr\"odinger-like wave equation with an effective potential that vanishes both at the event horizon and at spatial infinity in the black-hole regime. This common asymptotic structure makes the DBI-supported geometry a convenient setting for a direct spin-by-spin comparison of ringdown and wave propagation in a regular black-hole background.

The scalar barrier is highest, the electromagnetic one is purely centrifugal, and the Dirac peak lies closer to the horizon; increasing $a$ lowers and broadens all barriers.

The numerical results confirm this expectation. For the fundamental modes that admit a WKB treatment, increasing the DBI regularity parameter shifts the quasinormal frequencies toward the origin of the complex plane, so that both $\re{\omega}$ and $|\im{\omega}|$ decrease. The effect is moderate in absolute terms but clearly resolved numerically: over the studied interval, the shifts are typically of order $10\%$ for the representative scalar, electromagnetic, and Dirac modes discussed in Sec.~VI, whereas the quality factor changes only weakly. The scalar sector becomes slightly less oscillatory as $a$ increases, while in the electromagnetic and Dirac sectors the quality factor grows mildly. Therefore, the DBI regularity scale leaves a consistent imprint on the ringing, but it does so without qualitatively altering the single-barrier character of the problem.

An important outcome of the present analysis is that the observed spectral changes are significantly larger than the estimated numerical uncertainty. The internal accuracy test based on the difference between the WKB16 and WKB14 Pad\'e-improved values shows excellent stability for the fundamental modes and acceptable stability even for the higher overtones considered here. In most cases the discrepancy is many orders of magnitude smaller than the cumulative shift produced by changing $a$, and even in the least favorable cases it remains well below the physical effect. This conclusion is reinforced by the time-domain profile for the scalar mode, whose Prony fit agrees with the WKB result to better than $6\times 10^{-4}\%$. Thus, within the domain of applicability of the WKB method, the deviation from the Schwarzschild spectrum can be regarded as a robust feature of the DBI-supported regular black hole rather than a numerical artifact.

Taken together, these results establish the analytic and numerical groundwork for a broader perturbative study of the DBI geometry. The natural continuation is to enlarge the time-domain analysis, include a more systematic study of higher overtones and late-time tails, and exploit the single-barrier structure of the effective potentials for the computation of grey-body factors and related scattering characteristics. It would also be interesting to extend the comparison beyond the black-hole branch, toward the extremal-remnant and horizonless regular configurations of the same solution, and ultimately to include gravitational perturbations in order to determine which features of the ringdown are specific to the test fields considered here and which are genuine signatures of the underlying geometry.

\vspace{5mm}

\section*{Declaration of Competing Interest}
The authors declare that they have no known competing financial interests or personal relationships that could have appeared to influence the work reported in this paper.

\begin{acknowledgments}
B. C. L. is grateful to the Excellence project FoS UHK 2205/2025-2026 for the financial support.
\end{acknowledgments}

\section*{Data Availability}
No data was used for the research described in the article.

\bibliographystyle{apsrev4-1}
\bibliography{bibliography_massless}

@article{Karmakar:2023cwg,
    author = "Karmakar, Ronit and Goswami, Umananda Dev",
    title = "{Quasinormal modes, temperatures and greybody factors of black holes in a generalized Rastall gravity theory}",
    eprint = "2310.18594",
    archivePrefix = "arXiv",
    primaryClass = "gr-qc",
    doi = "10.1088/1402-4896/ad350e",
    journal = "Phys. Scripta",
    volume = "99",
    number = "5",
    pages = "055003",
    year = "2024"
}

@article{Matyjasek:2026yiu,
    author = "Matyjasek, Jerzy and Konoplya, Roman A. and Zhidenko, Alexander",
    title = "{An Efficient Higher-Order WKB Code for Quasinormal Modes and Greybody Factors}",
    eprint = "2603.12466",
    archivePrefix = "arXiv",
    primaryClass = "gr-qc",
    doi = "10.53941/ijgtp.2026.100005",
    journal = "Int. J. Grav. Theor. Phys.",
    volume = "2",
    number = "1",
    pages = "5",
    year = "2026"
}

@article{Konoplya:2017zwo,
    author = "Konoplya, R. A. and Zhidenko, A.",
    title = "{Quasinormal modes of Gauss-Bonnet-AdS black holes: towards holographic description of finite coupling}",
    eprint = "1705.07732",
    archivePrefix = "arXiv",
    primaryClass = "hep-th",
    doi = "10.1007/JHEP09(2017)139",
    journal = "JHEP",
    volume = "09",
    pages = "139",
    year = "2017"
}

@article{Konoplya:2024vuj,
    author = "Konoplya, R. A. and Zhidenko, A.",
    title = "{Correspondence between grey-body factors and quasinormal frequencies for rotating black holes}",
    eprint = "2408.11162",
    archivePrefix = "arXiv",
    primaryClass = "gr-qc",
    doi = "10.1016/j.physletb.2025.139288",
    journal = "Phys. Lett. B",
    volume = "861",
    pages = "139288",
    year = "2025"
}

@article{Al-Badawi:2023lke,
    author = "Al-Badawi, Ahmad and Kraishan, Amani",
    title = "{Dirac perturbations of Hayward black hole with quintessence: Quasinormal modes and greybody factor}",
    doi = "10.1016/j.cjph.2023.10.048",
    journal = "Chin. J. Phys.",
    volume = "87",
    pages = "59--69",
    year = "2024"
}

@article{Berti:2009kk,
    author = "Berti, Emanuele and Cardoso, Vitor and Starinets, Andrei O.",
    title = "{Quasinormal modes of black holes and black branes}",
    eprint = "0905.2975",
    archivePrefix = "arXiv",
    primaryClass = "gr-qc",
    doi = "10.1088/0264-9381/26/16/163001",
    journal = "Class. Quant. Grav.",
    volume = "26",
    pages = "163001",
    year = "2009"
}

@article{Bolokhov:2023bwm,
    author = "Bolokhov, S. V.",
    title = "{Long-lived quasinormal modes and overtones{\textquoteright} behavior of holonomy-corrected black holes}",
    eprint = "2311.05503",
    archivePrefix = "arXiv",
    primaryClass = "gr-qc",
    doi = "10.1103/PhysRevD.110.024010",
    journal = "Phys. Rev. D",
    volume = "110",
    number = "2",
    pages = "024010",
    year = "2024"
}

@article{Bolokhov:2026dzn,
    author = "Bolokhov, S. V.",
    title = "{Quasi-resonances in the vicinity of Einstein-Maxwell-dilaton black hole}",
    eprint = "2604.11845",
    archivePrefix = "arXiv",
    primaryClass = "gr-qc",
    month = "4",
    year = "2026"
}

@article{Breton:2017hwe,
    author = "Breton, Nora and Clark, Tyler and Fernando, Sharmanthie",
    title = "{Quasinormal modes and absorption cross-sections of Born-Infeld-de Sitter black holes}",
    eprint = "1703.10070",
    archivePrefix = "arXiv",
    primaryClass = "gr-qc",
    doi = "10.1142/S0218271817501127",
    journal = "Int. J. Mod. Phys. D",
    volume = "26",
    number = "10",
    pages = "1750112",
    year = "2017"
}

@article{Cai:2021ele,
    author = "Cai, Xin-Chang and Miao, Yan-Gang",
    title = "{Quasinormal modes and shadows of a new family of Ayón-Beato-García black holes}",
    eprint = "2104.09725",
    archivePrefix = "arXiv",
    primaryClass = "gr-qc",
    doi = "10.1103/PhysRevD.103.124050",
    journal = "Phys. Rev. D",
    volume = "103",
    number = "12",
    pages = "124050",
    year = "2021"
}

@article{DuttaRoy:2022ytr,
    author = "Dutta Roy, Poulami and Kar, Sayan",
    title = "{Generalized Hayward spacetimes: Geometry, matter, and scalar quasinormal modes}",
    eprint = "2206.04505",
    archivePrefix = "arXiv",
    primaryClass = "gr-qc",
    doi = "10.1103/PhysRevD.106.044028",
    journal = "Phys. Rev. D",
    volume = "106",
    number = "4",
    pages = "044028",
    year = "2022"
}

@article{Fernando:2012yw,
    author = "Fernando, Sharmanthie and Correa, Juan",
    title = "{Quasinormal Modes of Bardeen Black Hole: Scalar Perturbations}",
    eprint = "1208.5442",
    archivePrefix = "arXiv",
    primaryClass = "gr-qc",
    doi = "10.1103/PhysRevD.86.064039",
    journal = "Phys. Rev. D",
    volume = "86",
    pages = "064039",
    year = "2012"
}

@article{Flachi:2012nv,
    author = "Flachi, Antonino and Lemos, Jose P. S.",
    title = "{Quasinormal modes of regular black holes}",
    eprint = "1211.6212",
    archivePrefix = "arXiv",
    primaryClass = "gr-qc",
    doi = "10.1103/PhysRevD.87.024034",
    journal = "Phys. Rev. D",
    volume = "87",
    number = "2",
    pages = "024034",
    year = "2013"
}

@article{Gingrich:2024tuf,
    author = "Gingrich, Douglas M.",
    title = "{Quasinormal modes of a nonsingular spherically symmetric black hole effective model with holonomy corrections}",
    eprint = "2404.04447",
    archivePrefix = "arXiv",
    primaryClass = "gr-qc",
    doi = "10.1103/PhysRevD.110.084045",
    journal = "Phys. Rev. D",
    volume = "110",
    number = "8",
    pages = "084045",
    year = "2024"
}

@article{Gundlach:1993tp,
    author = "Gundlach, Carsten and Price, Richard H. and Pullin, Jorge",
    title = "{Late time behavior of stellar collapse and explosions: 1. Linearized perturbations}",
    eprint = "gr-qc/9307009",
    archivePrefix = "arXiv",
    reportNumber = "NSF-ITP-93-84",
    doi = "10.1103/PhysRevD.49.883",
    journal = "Phys. Rev. D",
    volume = "49",
    pages = "883--889",
    year = "1994"
}

@article{Guo:2024jhg,
    author = "Guo, Yang and Xie, Hao and Miao, Yan-Gang",
    title = "{Signal of phase transition hidden in quasinormal modes of regular AdS black holes}",
    eprint = "2402.10406",
    archivePrefix = "arXiv",
    primaryClass = "gr-qc",
    doi = "10.1016/j.physletb.2024.138801",
    journal = "Phys. Lett. B",
    volume = "855",
    pages = "138801",
    year = "2024"
}

@article{Held:2019xde,
    author = "Held, Aaron and Gold, Roman and Eichhorn, Astrid",
    title = "{Asymptotic safety casts its shadow}",
    eprint = "1904.07133",
    archivePrefix = "arXiv",
    primaryClass = "gr-qc",
    doi = "10.1088/1475-7516/2019/06/029",
    journal = "JCAP",
    volume = "06",
    pages = "029",
    year = "2019"
}

@article{Huang:2023aet,
    author = "Huang, Bai-Hao and Hu, Han-Wen and Zhao, Liu",
    title = "{Thermodynamics for regular black holes as intermediate thermodynamic states and quasinormal frequencies}",
    eprint = "2311.12286",
    archivePrefix = "arXiv",
    primaryClass = "gr-qc",
    doi = "10.1088/1475-7516/2024/03/053",
    journal = "JCAP",
    volume = "03",
    pages = "053",
    year = "2024"
}

@article{Ishihara:2008re,
    author = "Ishihara, Hideki and Kimura, Masashi and Konoplya, Roman A. and Murata, Keiju and Soda, Jiro and Zhidenko, Alexander",
    title = "{Evolution of perturbations of squashed Kaluza-Klein black holes: escape from instability}",
    eprint = "0802.0655",
    archivePrefix = "arXiv",
    primaryClass = "hep-th",
    doi = "10.1103/PhysRevD.77.084019",
    journal = "Phys. Rev. D",
    volume = "77",
    pages = "084019",
    year = "2008"
}

@article{Iyer:1986np,
    author = "Iyer, Sai and Will, Clifford M.",
    title = "{Black Hole Normal Modes: A {WKB} Approach. 1. Foundations and Application of a Higher Order {WKB} Analysis of Potential Barrier Scattering}",
    reportNumber = "Print-86-1482 (WASH. U., ST. LOUIS)",
    doi = "10.1103/PhysRevD.35.3621",
    journal = "Phys. Rev. D",
    volume = "35",
    pages = "3621",
    year = "1987"
}

@article{Jawad:2020hju,
    author = "Jawad, Abdul and Yasir, Muhammad and Rani, Shamaila",
    title = "{Joule–Thomson expansion and quasinormal modes of regular non-minimal magnetic black hole}",
    doi = "10.1142/S0217732320502983",
    journal = "Mod. Phys. Lett. A",
    volume = "35",
    number = "36",
    pages = "2050298",
    year = "2020"
}

@article{Jusufi:2020odz,
    author = {Jusufi, Kimet and Azreg-Ainou, Mustapha and Jamil, Mubasher and Wei, Shao-Wen and Wu, Qiang and Wang, Anzhong},
    title = "{Quasinormal modes, quasiperiodic oscillations, and the shadow of rotating regular black holes in nonminimally coupled Einstein-Yang-Mills theory}",
    eprint = "2008.08450",
    archivePrefix = "arXiv",
    primaryClass = "gr-qc",
    doi = "10.1103/PhysRevD.103.024013",
    journal = "Phys. Rev. D",
    volume = "103",
    number = "2",
    pages = "024013",
    year = "2021"
}

@article{Kokkotas:1999bd,
    author = "Kokkotas, Kostas D. and Schmidt, Bernd G.",
    title = "{Quasinormal modes of stars and black holes}",
    eprint = "gr-qc/9909058",
    archivePrefix = "arXiv",
    doi = "10.12942/lrr-1999-2",
    journal = "Living Rev. Rel.",
    volume = "2",
    pages = "2",
    year = "1999"
}

@article{Konoplya:2003ii,
    author = "Konoplya, R. A.",
    title = "{Quasinormal behavior of the d-dimensional Schwarzschild black hole and higher order WKB approach}",
    eprint = "gr-qc/0303052",
    archivePrefix = "arXiv",
    doi = "10.1103/PhysRevD.68.024018",
    journal = "Phys. Rev. D",
    volume = "68",
    pages = "024018",
    year = "2003"
}

@article{Konoplya:2010vz,
    author = "Konoplya, R. A. and Zhidenko, A.",
    title = "{Long life of Gauss-Bonnet corrected black holes}",
    eprint = "1004.3772",
    archivePrefix = "arXiv",
    primaryClass = "hep-th",
    doi = "10.1103/PhysRevD.82.084003",
    journal = "Phys. Rev. D",
    volume = "82",
    pages = "084003",
    year = "2010"
}

@article{Konoplya:2011qq,
    author = "Konoplya, R. A. and Zhidenko, A.",
    title = "{Quasinormal modes of black holes: From astrophysics to string theory}",
    eprint = "1102.4014",
    archivePrefix = "arXiv",
    primaryClass = "gr-qc",
    doi = "10.1103/RevModPhys.83.793",
    journal = "Rev. Mod. Phys.",
    volume = "83",
    pages = "793--836",
    year = "2011"
}

@article{Konoplya:2017lhs,
    author = "Konoplya, R. A. and Zhidenko, A.",
    title = "{The portrait of eikonal instability in Lovelock theories}",
    eprint = "1705.01656",
    archivePrefix = "arXiv",
    primaryClass = "hep-th",
    doi = "10.1088/1475-7516/2017/05/050",
    journal = "JCAP",
    volume = "05",
    pages = "050",
    year = "2017"
}

@article{Konoplya:2017ymp,
    author = "Konoplya, R. A. and Zhidenko, A.",
    title = "{Eikonal instability of Gauss-Bonnet–(anti-)–de Sitter black holes}",
    eprint = "1701.01652",
    archivePrefix = "arXiv",
    primaryClass = "hep-th",
    doi = "10.1103/PhysRevD.95.104005",
    journal = "Phys. Rev. D",
    volume = "95",
    number = "10",
    pages = "104005",
    year = "2017"
}

@article{Bolokhov:2023ruj,
    author = "Bolokhov, S. V.",
    title = "{Long-lived quasinormal modes and oscillatory tails of the Bardeen spacetime}",
    doi = "10.1103/PhysRevD.109.064017",
    journal = "Phys. Rev. D",
    volume = "109",
    number = "6",
    pages = "064017",
    year = "2024"
}

@article{Konoplya:2019ppy,
    author = "Konoplya, R. A. and Zinhailo, A. F.",
    title = "{Hawking radiation of non-Schwarzschild black holes in higher derivative gravity: a crucial role of grey-body factors}",
    eprint = "1904.05341",
    archivePrefix = "arXiv",
    primaryClass = "gr-qc",
    doi = "10.1103/PhysRevD.99.104060",
    journal = "Phys. Rev. D",
    volume = "99",
    number = "10",
    pages = "104060",
    year = "2019"
}

@article{Konoplya:2018ala,
    author = "Konoplya, R. A.",
    title = "{How to tell the shape of a wormhole by its quasinormal modes}",
    eprint = "1805.04718",
    archivePrefix = "arXiv",
    primaryClass = "gr-qc",
    doi = "10.1016/j.physletb.2018.07.025",
    journal = "Phys. Lett. B",
    volume = "784",
    pages = "43--49",
    year = "2018"
}

@article{Konoplya:2020pbh,
    author = "Konoplya, R. A. and Zhidenko, A.",
    title = "{General parametrization of black holes: The only parameters that matter}",
    eprint = "2001.06100",
    archivePrefix = "arXiv",
    primaryClass = "gr-qc",
    doi = "10.1103/PhysRevD.101.124004",
    journal = "Phys. Rev. D",
    volume = "101",
    number = "12",
    pages = "124004",
    year = "2020"
}

@article{Konoplya:2022hll,
    author = "Konoplya, R. A. and Zinhailo, A. F. and Kunz, J. and Stuchlik, Z. and Zhidenko, A.",
    title = "{Quasinormal ringing of regular black holes in asymptotically safe gravity: the importance of overtones}",
    eprint = "2206.14714",
    archivePrefix = "arXiv",
    primaryClass = "gr-qc",
    doi = "10.1088/1475-7516/2022/10/091",
    journal = "JCAP",
    volume = "10",
    pages = "091",
    year = "2022"
}

@article{Konoplya:2023fmh,
    author = "Konoplya, R. A. and Zhidenko, A.",
    title = "{Asymptotic tails of massive gravitons in light of pulsar timing array observations}",
    eprint = "2307.01110",
    archivePrefix = "arXiv",
    primaryClass = "gr-qc",
    doi = "10.1016/j.physletb.2024.138685",
    journal = "Phys. Lett. B",
    volume = "853",
    pages = "138685",
    year = "2024"
}

@article{Konoplya:2023ppx,
    author = "Konoplya, R. A.",
    title = "{Quasinormal modes and grey-body factors of regular black holes with a scalar hair from the Effective Field Theory}",
    eprint = "2305.09187",
    archivePrefix = "arXiv",
    primaryClass = "gr-qc",
    doi = "10.1088/1475-7516/2023/07/001",
    journal = "JCAP",
    volume = "07",
    pages = "001",
    year = "2023"
}

@article{Li:2014fka,
    author = "Li, Jin and Lin, Kai and Yang, Nan",
    title = "{Nonlinear electromagnetic quasinormal modes and Hawking radiation of a regular black hole with magnetic charge}",
    eprint = "1409.5988",
    archivePrefix = "arXiv",
    primaryClass = "gr-qc",
    doi = "10.1140/epjc/s10052-015-3347-3",
    journal = "Eur. Phys. J. C",
    volume = "75",
    number = "3",
    pages = "131",
    year = "2015"
}

@article{Lin:2013ofa,
    author = "Lin, Kai and Li, Jin and Yang, Shuzheng",
    title = "{Quasinormal Modes of Hayward Regular Black Hole}",
    doi = "10.1007/s10773-013-1682-4",
    journal = "Int. J. Theor. Phys.",
    volume = "52",
    pages = "3771--3778",
    year = "2013"
}

@article{Lopez:2022uie,
    author = "Lopez, L. A. and Ramírez, Valeria",
    title = "{Quasi-normal modes of a Generic-class of magnetically charged regular black hole: scalar and electromagnetic perturbations}",
    eprint = "2205.10166",
    archivePrefix = "arXiv",
    primaryClass = "gr-qc",
    doi = "10.1140/epjp/s13360-023-03735-6",
    journal = "Eur. Phys. J. Plus",
    volume = "138",
    number = "2",
    pages = "120",
    year = "2023"
}

@article{Lutfuoglu:2025eik,
    author = {L{ü}tf{ü}o{\u{g}}lu, Bekir Can},
    title = "{Grey-Body Factors and Absorption Cross-Sections~of Scalar and Dirac Fields in the Vicinity of Dilaton-De Sitter Black Hole}",
    eprint = "2510.10579",
    archivePrefix = "arXiv",
    primaryClass = "gr-qc",
    doi = "10.1002/prop.70074",
    journal = "Fortsch. Phys.",
    volume = "74",
    number = "1",
    pages = "e70074",
    year = "2026"
}

@article{Lutfuoglu:2025kqp,
    author = {L{ü}tf{ü}o{\u{g}}lu, B. C.},
    title = "{Long-lived quasinormal modes, grey-body factors and absorption cross-section of the black hole immersed in the Hernquist galactic halo}",
    eprint = "2510.25969",
    archivePrefix = "arXiv",
    primaryClass = "gr-qc",
    doi = "10.1016/j.physletb.2025.140082",
    journal = "Phys. Lett. B",
    volume = "872",
    pages = "140082",
    year = "2026"
}

@article{Lutfuoglu:2026uzy,
    author = {L{ü}tf{ü}o{\u{g}}lu, Bekir Can},
    title = "{Quasinormal Modes of a Massive Scalar Field in 4D Einstein--Gauss--Bonnet Black Hole Spacetimes}",
    eprint = "2603.24424",
    archivePrefix = "arXiv",
    primaryClass = "gr-qc",
    month = "3",
    year = "2026"
}

@article{MahdavianYekta:2019pol,
    author = "Mahdavian Yekta, Davood and Karimabadi, Majid and Alavi, S. A.",
    title = "{Quasinormal modes for non-minimally coupled scalar fields in regular black hole spacetimes: Grey-body factors, area spectrum and shadow radius}",
    eprint = "1912.12017",
    archivePrefix = "arXiv",
    primaryClass = "hep-th",
    doi = "10.1016/j.aop.2021.168603",
    journal = "Annals Phys.",
    volume = "434",
    pages = "168603",
    year = "2021"
}

@article{Matyjasek:2017psv,
    author = "Matyjasek, Jerzy and Opala, Michał",
    title = "{Quasinormal modes of black holes. The improved semianalytic approach}",
    eprint = "1704.00361",
    archivePrefix = "arXiv",
    primaryClass = "gr-qc",
    doi = "10.1103/PhysRevD.96.024011",
    journal = "Phys. Rev. D",
    volume = "96",
    number = "2",
    pages = "024011",
    year = "2017"
}

@article{Mukohyama:2023xyf,
    author = "Mukohyama, Shinji and Takahashi, Kazufumi and Tomikawa, Keitaro and Yingcharoenrat, Vicharit",
    title = "{Quasinormal modes from EFT of black hole perturbations with timelike scalar profile}",
    eprint = "2304.14304",
    archivePrefix = "arXiv",
    primaryClass = "gr-qc",
    reportNumber = "YITP-23-45, IPMU23-0007, RUP-23-8",
    doi = "10.1088/1475-7516/2023/07/050",
    journal = "JCAP",
    volume = "07",
    pages = "050",
    year = "2023"
}

@article{Pedraza:2021hzw,
    author = "Pedraza, Omar and López, L. A. and Arceo, R. and Cabrera-Munguia, I.",
    title = "{Quasinormal modes of the Hayward black hole surrounded by quintessence: Scalar, electromagnetic and gravitational perturbations}",
    eprint = "2111.06488",
    archivePrefix = "arXiv",
    primaryClass = "gr-qc",
    doi = "10.1142/S0217732322500572",
    journal = "Mod. Phys. Lett. A",
    volume = "37",
    number = "09",
    pages = "2250057",
    year = "2022"
}

@article{Bolokhov:2026kqu,
    author = "Bolokhov, S. V.",
    title = "{Scattering of scalar, electromagnetic, and Dirac fields in an asymptotically flat regular black hole supported by primordial dark matter}",
    eprint = "2605.03137",
    archivePrefix = "arXiv",
    primaryClass = "gr-qc",
    month = "5",
    year = "2026"
}

@article{Parvez:2025dbi,
    author = "Parvez, Tausif and Shankaranarayanan, S.",
    title = "{Exact, nonsingular black holes from a phantom DBI field as primordial dark matter}",
    eprint = "2511.14047",
    archivePrefix = "arXiv",
    primaryClass = "gr-qc",
    doi = "10.1103/118s-qpgy",
    journal = "Phys. Rev. D",
    volume = "113",
    number = "10",
    pages = "L101503",
    year = "2026"
}

@article{Skvortsova:2026unq,
    author = "Skvortsova, Milena",
    title = "{Long-lived quasinormal frequencies for regular black hole supported by the Einasto profile in the presence of the magnetic field}",
    eprint = "2603.28415",
    archivePrefix = "arXiv",
    primaryClass = "gr-qc",
    month = "3",
    year = "2026"
}

@article{Wongjun:2019ydo,
    author = "Wongjun, Pitayuth and Chen, Chun-Hung and Nakarachinda, Ratchaphat",
    title = "{Quasinormal modes of a massless Dirac field in de Rham-Gabadadze-Tolley massive gravity}",
    eprint = "1910.05908",
    archivePrefix = "arXiv",
    primaryClass = "gr-qc",
    doi = "10.1103/PhysRevD.101.124033",
    journal = "Phys. Rev. D",
    volume = "101",
    number = "12",
    pages = "124033",
    year = "2020"
}

@article{Yang:2021cvh,
    author = "Yang, Yi and Liu, Dong and Xu, Zhaoyi and Xing, Yujia and Wu, Shurui and Long, Zheng-Wen",
    title = "{Echoes of novel black-bounce spacetimes}",
    eprint = "2107.06554",
    archivePrefix = "arXiv",
    primaryClass = "gr-qc",
    doi = "10.1103/PhysRevD.104.104021",
    journal = "Phys. Rev. D",
    volume = "104",
    number = "10",
    pages = "104021",
    year = "2021"
}

@article{Zhang:2024nny,
    author = "Zhang, Dan and Gong, Huajie and Fu, Guoyang and Wu, Jian-Pin and Pan, Qiyuan",
    title = "{Quasinormal modes of a regular black hole with sub-Planckian curvature}",
    eprint = "2402.15085",
    archivePrefix = "arXiv",
    primaryClass = "gr-qc",
    doi = "10.1140/epjc/s10052-024-12928-x",
    journal = "Eur. Phys. J. C",
    volume = "84",
    number = "6",
    pages = "564",
    year = "2024"
}

@article{Konoplya:2024gbf,
    author = "Konoplya, R. A. and Zhidenko, A.",
    title = "{Correspondence between grey-body factors and quasinormal modes}",
    eprint = "2406.11694",
    archivePrefix = "arXiv",
    primaryClass = "gr-qc",
    doi = "10.1088/1475-7516/2024/09/068",
    journal = "JCAP",
    volume = "09",
    pages = "068",
    year = "2024"
}

@article{Dubinsky:2025fwv,
    author = "Dubinsky, Alexey",
    title = "{Black Holes Immersed in Galactic Dark Matter Halo}",
    eprint = "2507.00256",
    archivePrefix = "arXiv",
    primaryClass = "gr-qc",
    doi = "10.53941/ijgtp.2025.100002",
    journal = "Int. J. Grav. Theor. Phys.",
    volume = "1",
    number = "1",
    pages = "2",
    year = "2025"
}

@article{Bolokhov:2024ixe,
    author = "Bolokhov, S. V.",
    title = "{Late time decay of scalar and Dirac fields around an asymptotically de Sitter black hole in the Euler{\textendash}Heisenberg electrodynamics}",
    eprint = "2404.09364",
    archivePrefix = "arXiv",
    primaryClass = "gr-qc",
    doi = "10.1140/epjc/s10052-024-12990-5",
    journal = "Eur. Phys. J. C",
    volume = "84",
    number = "6",
    pages = "634",
    year = "2024"
}

@article{Malik:2024nhy,
    author = "Malik, Zainab",
    title = "{Perturbations and quasinormal modes of the Dirac field in Effective Quantum Gravity}",
    eprint = "2409.01561",
    archivePrefix = "arXiv",
    primaryClass = "gr-qc",
    doi = "10.1016/j.aop.2025.170046",
    journal = "Annals Phys.",
    volume = "479",
    pages = "170046",
    year = "2025"
}

@article{Konoplya:2024hfg,
    author = "Konoplya, R. A. and Zhidenko, A.",
    title = "{Infinite tower of higher-curvature corrections: Quasinormal modes and late-time behavior of D-dimensional regular black holes}",
    eprint = "2403.07848",
    archivePrefix = "arXiv",
    primaryClass = "gr-qc",
    doi = "10.1103/PhysRevD.109.104005",
    journal = "Phys. Rev. D",
    volume = "109",
    number = "10",
    pages = "104005",
    year = "2024"
}

@article{Bolokhov:2024otn,
    author = "Bolokhov, S. V. and Skvortsova, Milena",
    title = "{Correspondence between quasinormal modes and grey-body factors of spherically symmetric traversable wormholes}",
    eprint = "2412.11166",
    archivePrefix = "arXiv",
    primaryClass = "gr-qc",
    doi = "10.1088/1475-7516/2025/04/025",
    journal = "JCAP",
    volume = "04",
    pages = "025",
    year = "2025"
}

@article{Skvortsova:2024atk,
    author = "Skvortsova, Milena",
    title = "{Quasinormal Frequencies of Fields with Various Spin in the Quantum Oppenheimer{\textendash}Snyder Model of Black Holes}",
    eprint = "2405.06390",
    archivePrefix = "arXiv",
    primaryClass = "gr-qc",
    doi = "10.1002/prop.202400132",
    journal = "Fortsch. Phys.",
    volume = "72",
    number = "9-10",
    pages = "2400132",
    year = "2024"
}

@article{Malik:2023bxc,
    author = "Malik, Zainab",
    title = "{Quasinormal Modes of the Bumblebee Black Holes with a Global Monopole}",
    eprint = "2308.10412",
    archivePrefix = "arXiv",
    primaryClass = "gr-qc",
    doi = "10.1007/s10773-024-05737-1",
    journal = "Int. J. Theor. Phys.",
    volume = "63",
    number = "8",
    pages = "199",
    year = "2024"
}

@article{Varghese:2011ku,
    author = "Varghese, Nijo and Kuriakose, V. C.",
    title = "{Evolution of massive fields around a black hole in Horava gravity}",
    eprint = "1011.6608",
    archivePrefix = "arXiv",
    primaryClass = "gr-qc",
    doi = "10.1007/s10714-011-1201-y",
    journal = "Gen. Rel. Grav.",
    volume = "43",
    pages = "2757--2767",
    year = "2011"
}

@article{Skvortsova:2023zmj,
    author = "Skvortsova, Milena",
    title = "{Quasinormal Spectrum of (2+1)-Dimensional Asymptotically Flat, dS and AdS Black Holes}",
    eprint = "2311.11650",
    archivePrefix = "arXiv",
    primaryClass = "gr-qc",
    doi = "10.1002/prop.202400036",
    journal = "Fortsch. Phys.",
    volume = "72",
    number = "6",
    pages = "2400036",
    year = "2024"
}

@article{Skvortsova:2024wly,
    author = "Skvortsova, Milena",
    title = "{Ringing of Extreme Regular Black Holes}",
    eprint = "2405.15807",
    archivePrefix = "arXiv",
    primaryClass = "gr-qc",
    doi = "10.1134/S020228932470018X",
    journal = "Grav. Cosmol.",
    volume = "30",
    number = "3",
    pages = "279--288",
    year = "2024"
}

@article{Lutfuoglu:2025hjy,
    author = {L{\"u}tf{\"u}o{\u{g}}lu, B. C.},
    title = "{Long-lived quasinormal modes and gray-body factors of black holes and wormholes in dark matter inspired Weyl gravity}",
    eprint = "2503.16087",
    archivePrefix = "arXiv",
    primaryClass = "gr-qc",
    doi = "10.1140/epjc/s10052-025-14210-0",
    journal = "Eur. Phys. J. C",
    volume = "85",
    number = "5",
    pages = "486",
    year = "2025"
}

@article{Bolokhov:2023dxq,
    author = "Bolokhov, S. V.",
    title = "{Black holes in Starobinsky-Bel-Robinson Gravity and the breakdown of quasinormal modes/null geodesics correspondence}",
    eprint = "2310.12326",
    archivePrefix = "arXiv",
    primaryClass = "gr-qc",
    doi = "10.1016/j.physletb.2024.138879",
    journal = "Phys. Lett. B",
    volume = "856",
    pages = "138879",
    year = "2024"
}

@article{Dubinsky:2024nzo,
    author = "Dubinsky, Alexey and Zinhailo, Antonina F.",
    title = "{Analytic expressions for grey-body factors of the general parametrized spherically symmetric black holes}",
    eprint = "2410.15232",
    archivePrefix = "arXiv",
    primaryClass = "gr-qc",
    doi = "10.1209/0295-5075/adbc17",
    journal = "EPL",
    volume = "149",
    number = "6",
    pages = "69004",
    year = "2025"
}

@article{Abdalla:2012si,
    author = "Abdalla, Elcio and Piedra, Owen Pavel Fernandez and Nu{\~n}ez, Fidel Sosa and de Oliveira, Jeferson",
    title = "{Scalar field propagation in higher dimensional black holes at a Lifshitz point}",
    eprint = "1211.3390",
    archivePrefix = "arXiv",
    primaryClass = "gr-qc",
    reportNumber = "GEA-UCF-2012-03",
    doi = "10.1103/PhysRevD.88.064035",
    journal = "Phys. Rev. D",
    volume = "88",
    number = "6",
    pages = "064035",
    year = "2013"
}

@article{Bolokhov:2026eqf,
    author = "Bolokhov, S. V.",
    title = "{Quasinormal Modes and Grey-Body Factors of Scalar, Electromagnetic and Dirac Fields Around Einasto-Supported Regular Black Holes}",
    eprint = "2603.22310",
    archivePrefix = "arXiv",
    primaryClass = "gr-qc",
    month = "3",
    year = "2026"
}

@article{Arbelaez:2026eaz,
    author = "Arbelaez, Juan Pablo",
    title = "{Grey-body factors of higher dimensional regular black holes in quasi-topological theories}",
    eprint = "2601.22340",
    archivePrefix = "arXiv",
    primaryClass = "gr-qc",
    month = "1",
    year = "2026"
}

@article{Arbelaez:2025gwj,
    author = "Arbelaez, Juan Pablo",
    title = "{Quasinormal spectra of higher dimensional regular black holes in theories with infinite curvature corrections}",
    eprint = "2509.25141",
    archivePrefix = "arXiv",
    primaryClass = "gr-qc",
    month = "9",
    year = "2025"
}

@article{Abdalla:2005hu,
    author = "Abdalla, E. and Konoplya, R. A. and Molina, C.",
    title = "{Scalar field evolution in Gauss-Bonnet black holes}",
    eprint = "hep-th/0507100",
    archivePrefix = "arXiv",
    doi = "10.1103/PhysRevD.72.084006",
    journal = "Phys. Rev. D",
    volume = "72",
    pages = "084006",
    year = "2005"
}

@article{Fernando:2016ftj,
    author = "Fernando, Sharmanthie",
    title = "{Quasinormal modes of dilaton-de Sitter black holes: scalar perturbations}",
    eprint = "1601.06407",
    archivePrefix = "arXiv",
    primaryClass = "gr-qc",
    doi = "10.1007/s10714-016-2020-y",
    journal = "Gen. Rel. Grav.",
    volume = "48",
    number = "3",
    pages = "24",
    year = "2016"
}

@article{Bolokhov:2025egl,
    author = "Bolokhov, S. V. and Skvortsova, Milena",
    title = "{Gravitational quasinormal modes of the Hayward spacetime}",
    eprint = "2508.19989",
    archivePrefix = "arXiv",
    primaryClass = "gr-qc",
    doi = "10.1140/epjc/s10052-026-15624-0",
    journal = "Eur. Phys. J. C",
    volume = "86",
    number = "4",
    pages = "374",
    year = "2026"
}

@article{Konoplya:2005sy,
    author = "Konoplya, R. A. and Abdalla, Elcio",
    title = "{Scalar field perturbations of the Schwarzschild black hole in the Godel universe}",
    eprint = "hep-th/0503029",
    archivePrefix = "arXiv",
    doi = "10.1103/PhysRevD.71.084015",
    journal = "Phys. Rev. D",
    volume = "71",
    pages = "084015",
    year = "2005"
}

@article{Konoplya:2001ji,
    author = "Konoplya, R. A.",
    title = "{Quasinormal modes of the electrically charged dilaton black hole}",
    eprint = "gr-qc/0109096",
    archivePrefix = "arXiv",
    doi = "10.1023/A:1015347628961",
    journal = "Gen. Rel. Grav.",
    volume = "34",
    pages = "329--335",
    year = "2002"
}

@article{Konoplya:2006ar,
    author = "Konoplya, R. A. and Zhidenko, A.",
    title = "{Gravitational spectrum of black holes in the Einstein-Aether theory}",
    eprint = "hep-th/0611226",
    archivePrefix = "arXiv",
    doi = "10.1016/j.physletb.2007.03.018",
    journal = "Phys. Lett. B",
    volume = "648",
    pages = "236--239",
    year = "2007"
}

@article{Kokkotas:2010zd,
    author = "Kokkotas, K. D. and Konoplya, R. A. and Zhidenko, A.",
    title = "{Quasinormal modes, scattering and Hawking radiation of Kerr-Newman black holes in a magnetic field}",
    eprint = "1011.1843",
    archivePrefix = "arXiv",
    primaryClass = "gr-qc",
    doi = "10.1103/PhysRevD.83.024031",
    journal = "Phys. Rev. D",
    volume = "83",
    pages = "024031",
    year = "2011"
}

@article{Malik:2025erb,
    author = "Malik, Zainab",
    title = "{Grey-Body Factors for Scalar and Dirac Fields in the Euler-Heisenberg Electrodynamics}",
    eprint = "2509.15995",
    archivePrefix = "arXiv",
    primaryClass = "gr-qc",
    doi = "10.53941/ijgtp.2025.100006",
    journal = "Int. J. Grav. Theor. Phys.",
    volume = "1",
    number = "1",
    pages = "6",
    year = "2025"
}

@article{Pathrikar:2025gzu,
    author = "Pathrikar, Akshat",
    title = "{Quasinormal Ringing and Unruh-Verlinde Temperature of the Frolov Black Hole}",
    eprint = "2510.01376",
    archivePrefix = "arXiv",
    primaryClass = "gr-qc",
    doi = "10.53941/ijgtp.2026.100001",
    journal = "Int. J. Grav. Theor. Phys.",
    volume = "1",
    number = "1",
    pages = "1",
    year = "2026"
}

@article{Momennia:2018hsm,
    author = "Momennia, Mehrab and Hossein Hendi, Seyed and Soltani Bidgoli, Fatemeh",
    title = "{Stability and quasinormal modes of black holes in conformal Weyl gravity}",
    eprint = "1807.01792",
    archivePrefix = "arXiv",
    primaryClass = "hep-th",
    doi = "10.1016/j.physletb.2020.136028",
    journal = "Phys. Lett. B",
    volume = "813",
    pages = "136028",
    year = "2021"
}

@article{Guo:2020caw,
    author = "Guo, Yang and Miao, Yan-Gang",
    title = "{Scalar quasinormal modes of black holes in Einstein-Yang-Mills gravity}",
    eprint = "2005.07524",
    archivePrefix = "arXiv",
    primaryClass = "hep-th",
    doi = "10.1103/PhysRevD.102.064049",
    journal = "Phys. Rev. D",
    volume = "102",
    number = "6",
    pages = "064049",
    year = "2020"
}

@article{Konoplya:2009hv,
    author = "Konoplya, R. A. and Zhidenko, A.",
    title = "{Holographic conductivity of zero temperature superconductors}",
    eprint = "0909.2138",
    archivePrefix = "arXiv",
    primaryClass = "hep-th",
    doi = "10.1016/j.physletb.2010.02.048",
    journal = "Phys. Lett. B",
    volume = "686",
    pages = "199--206",
    year = "2010"
}

@article{Eniceicu:2019npi,
    author = "Eniceicu, Dan Stefan and Reece, Matthew",
    title = {{Quasinormal modes of charged fields in Reissner-Nordstr{\"o}m backgrounds by Borel-Pad{\'e} summation of Bender-Wu series}},
    eprint = "1912.05553",
    archivePrefix = "arXiv",
    primaryClass = "gr-qc",
    doi = "10.1103/PhysRevD.102.044015",
    journal = "Phys. Rev. D",
    volume = "102",
    number = "4",
    pages = "044015",
    year = "2020"
}

@article{Bolokhov:2025aqy,
    author = "Bolokhov, Sergei V. and Bronnikov, Kirill A. and Skvortsova, Milena V.",
    title = "{Stability Ranges of Magnetic Black Holes and Mirror (Topological) Stars in 5D Gravity}",
    eprint = "2507.14603",
    archivePrefix = "arXiv",
    primaryClass = "gr-qc",
    doi = "10.53941/ijgtp.2026.100002",
    journal = "Int. J. Grav. Theor. Phys.",
    volume = "2",
    number = "1",
    pages = "2",
    year = "2026"
}

@article{Bolokhov:2025lnt,
    author = "Bolokhov, S. V. and Skvortsova, Milena",
    title = "{Gravitational Quasinormal Modes and Grey-Body Factors of Bonanno{\textendash}Reuter Regular Black Holes}",
    eprint = "2507.07196",
    archivePrefix = "arXiv",
    primaryClass = "gr-qc",
    doi = "10.53941/ijgtp.2025.100003",
    journal = "Int. J. Grav. Theor. Phys.",
    volume = "1",
    number = "1",
    pages = "3",
    year = "2025"
}

@article{Skvortsova:2024msa,
    author = "Skvortsova, Milena",
    title = "{Quantum corrected black holes: testing the correspondence between grey-body factors and quasinormal modes}",
    eprint = "2411.06007",
    archivePrefix = "arXiv",
    primaryClass = "gr-qc",
    doi = "10.1140/epjc/s10052-025-14589-w",
    journal = "Eur. Phys. J. C",
    volume = "85",
    number = "8",
    pages = "854",
    year = "2025"
}

@article{Malik:2026lfj,
    author = "Malik, Zainab",
    title = "{Analytic Expressions for Quasinormal Modes of a Regular Black Hole Sourced by a Dehnen-Type Halo}",
    eprint = "2603.18887",
    archivePrefix = "arXiv",
    primaryClass = "gr-qc",
    doi = "10.53941/ijgtp.2026.100003",
    journal = "Int. J. Grav. Theor. Phys.",
    volume = "2",
    number = "1",
    pages = "3",
    year = "2026"
}

@article{Konoplya:2013sba,
    author = "Konoplya, R. A. and Zhidenko, A.",
    title = "{Instability of D-dimensional extremally charged Reissner-Nordstrom(-de Sitter) black holes: Extrapolation to arbitrary D}",
    eprint = "1309.7667",
    archivePrefix = "arXiv",
    primaryClass = "hep-th",
    doi = "10.1103/PhysRevD.89.024011",
    journal = "Phys. Rev. D",
    volume = "89",
    number = "2",
    pages = "024011",
    year = "2014"
}

@article{Takahashi:2010gz,
    author = "Takahashi, Tomohiro and Soda, Jiro",
    title = "{Catastrophic Instability of Small Lovelock Black Holes}",
    eprint = "1008.1618",
    archivePrefix = "arXiv",
    primaryClass = "gr-qc",
    reportNumber = "KUNS-2288",
    doi = "10.1143/PTP.124.711",
    journal = "Prog. Theor. Phys.",
    volume = "124",
    pages = "711--729",
    year = "2010"
}

@article{Dotti:2004sh,
    author = "Dotti, Gustavo and Gleiser, Reinaldo J.",
    title = "{Gravitational instability of Einstein-Gauss-Bonnet black holes under tensor mode perturbations}",
    eprint = "gr-qc/0409005",
    archivePrefix = "arXiv",
    doi = "10.1088/0264-9381/22/1/L01",
    journal = "Class. Quant. Grav.",
    volume = "22",
    pages = "L1",
    year = "2005"
}

@article{Konoplya:2025hgp,
    author = "Konoplya, R. A. and Stashko, O. S.",
    title = "{Transition from regular black holes to wormholes in covariant effective quantum gravity: Scattering, quasinormal modes, and Hawking radiation}",
    eprint = "2502.05689",
    archivePrefix = "arXiv",
    primaryClass = "gr-qc",
    doi = "10.1103/PhysRevD.111.084031",
    journal = "Phys. Rev. D",
    volume = "111",
    number = "8",
    pages = "084031",
    year = "2025"
}

@article{Lutfuoglu:2025mqa,
    author = {L{\"u}tf{\"u}o{\u{g}}lu, Bekir Can and Shermatov, Abubakir and Rayimbaev, Javlon and Matyoqubov, Muhammad and Sirajiddin, Otaboyev},
    title = "{Gravitational spectra and wave propagation in regular black holes supported by a Dehnen Halo}",
    eprint = "2511.22366",
    archivePrefix = "arXiv",
    primaryClass = "gr-qc",
    doi = "10.1140/epjc/s10052-025-15234-2",
    journal = "Eur. Phys. J. C",
    volume = "85",
    number = "12",
    pages = "1484",
    year = "2025"
}

@article{Lutfuoglu:2026gey,
    author = {L{\"u}tf{\"u}o{\u{g}}lu, Bekir Can and Rayimbaev, Javlon and Murodov, Sardor and Kurbanov, Jakhongir and Matyoqubov, Muhammad},
    title = "{A First-Order Eikonal Framework for Quasinormal Modes, Shadows, Strong Lensing, and Grey-Body Factors in a Scalarized Black-Hole Metric}",
    eprint = "2604.14999",
    archivePrefix = "arXiv",
    primaryClass = "gr-qc",
    doi = "10.1016/j.aop.2026.170514",
    journal = "Annals Phys.",
    volume = "491",
    pages = "170514",
    year = "2026"
}

@article{Konoplya:2023aph,
    author = "Konoplya, R. A. and Stuchlik, Z. and Zhidenko, A. and Zinhailo, A. F.",
    title = "{Quasinormal modes of renormalization group improved Dymnikova regular black holes}",
    eprint = "2303.01987",
    archivePrefix = "arXiv",
    primaryClass = "gr-qc",
    doi = "10.1103/PhysRevD.107.104050",
    journal = "Phys. Rev. D",
    volume = "107",
    number = "10",
    pages = "104050",
    year = "2023"
}

@article{Konoplya:2025ect,
    author = "Konoplya, R. A. and Zhidenko, A.",
    title = "{Dark matter halo as a source of regular black-hole geometries}",
    eprint = "2511.03066",
    archivePrefix = "arXiv",
    primaryClass = "gr-qc",
    doi = "10.1103/7ptp-9j1t",
    journal = "Phys. Rev. D",
    volume = "113",
    number = "4",
    pages = "043011",
    year = "2026"
}

@article{Dubinsky:2024vbn,
    author = "Dubinsky, Alexey",
    title = "{Gray-body factors for gravitational and electromagnetic perturbations around Gibbons{\textendash}Maeda{\textendash}Garfinkle{\textendash}Horowitz{\textendash}Strominger black holes}",
    eprint = "2412.00625",
    archivePrefix = "arXiv",
    primaryClass = "gr-qc",
    doi = "10.1142/S0217732325501111",
    journal = "Mod. Phys. Lett. A",
    volume = "40",
    number = "28",
    pages = "2550111",
    year = "2025"
}

@article{Malik:2025qnr,
    author = "Malik, Zainab",
    title = "{Bonanno-Reuter regular black hole: quasi-resonances, grey-body factors and absorption cross-sections of a massive scalar field}",
    eprint = "2510.06689",
    archivePrefix = "arXiv",
    primaryClass = "gr-qc",
    month = "10",
    year = "2025"
}

@article{Malik:2024wvs,
    author = "Malik, Zainab",
    title = "{Quasinormal modes and grey-body factors of Morris-Thorne wormholes}",
    eprint = "2412.13385",
    archivePrefix = "arXiv",
    primaryClass = "gr-qc",
    month = "12",
    year = "2024"
}

@article{Konoplya:2026gim,
    author = "Konoplya, R. A.",
    title = "{Quasinormal modes of four-dimensional regular black holes in quasi-topological gravity: Overtones{\textquoteright} outburst via WKB method}",
    eprint = "2603.03189",
    archivePrefix = "arXiv",
    primaryClass = "gr-qc",
    doi = "10.1016/j.physletb.2026.140386",
    journal = "Phys. Lett. B",
    volume = "876",
    pages = "140386",
    year = "2026"
}

@article{Bolokhov:2025uxz,
    author = "Bolokhov, S. V. and Skvortsova, Milena",
    title = "{Review of analytic results on quasinormal modes of black holes}",
    doi = "10.1134/S0202289325700306",
    journal = "Grav. Cosmol.",
    volume = "31",
    number = "4",
    pages = "423--446",
    eprint = "2504.05014",
    archivePrefix = "arXiv",
    primaryClass = "gr-qc",
    year = "2025"
}

@article{Saka:2025xxl,
    author = "Saka, Erdin{\c{c}} Ula{\c{s}}",
    title = "{Regular black hole sourced by the Dehnen-type distribution of matter: The sound of the event horizon}",
    eprint = "2512.08904",
    archivePrefix = "arXiv",
    primaryClass = "gr-qc",
    month = "12",
    year = "2025"
}

@article{Lutfuoglu:2025ljm,
    author = {L{\"u}tf{\"u}o{\u{g}}lu, B. C.},
    title = "{Non-minimal Einstein{\textendash}Yang{\textendash}Mills black holes: fundamental quasinormal mode and grey-body factors versus outburst of overtones}",
    eprint = "2504.18482",
    archivePrefix = "arXiv",
    primaryClass = "gr-qc",
    doi = "10.1140/epjc/s10052-025-14380-x",
    journal = "Eur. Phys. J. C",
    volume = "85",
    number = "6",
    pages = "630",
    year = "2025"
}

@article{Konoplya:2023ahd,
    author = "Konoplya, R. A. and Ovchinnikov, D. and Ahmedov, B.",
    title = "{Bardeen spacetime as a quantum corrected Schwarzschild black hole: Quasinormal modes and Hawking radiation}",
    eprint = "2307.10801",
    archivePrefix = "arXiv",
    primaryClass = "gr-qc",
    doi = "10.1103/PhysRevD.108.104054",
    journal = "Phys. Rev. D",
    volume = "108",
    number = "10",
    pages = "104054",
    year = "2023"
}

@article{Lutfuoglu:2025ldc,
    author = {L{\"u}tf{\"u}o{\u{g}}lu, Bekir Can},
    title = "{Black Holes in Proca-Gauss-Bonnet Gravity with Primary Hair: Particle Motion, Shadows, and Grey-Body Factors}",
    eprint = "2507.09246",
    archivePrefix = "arXiv",
    primaryClass = "gr-qc",
    doi = "10.53941/ijgtp.2025.100004",
    journal = "Int. J. Grav. Theor. Phys.",
    volume = "1",
    number = "1",
    pages = "4",
    year = "2025"
}

@article{Fernando:2003wc,
    author = "Fernando, Sharmanthie and Arnold, Keith",
    title = "{Scalar perturbations of charged dilaton black holes}",
    eprint = "hep-th/0312041",
    archivePrefix = "arXiv",
    reportNumber = "NKU-03-SF3",
    doi = "10.1023/B:GERG.0000035953.31652.88",
    journal = "Gen. Rel. Grav.",
    volume = "36",
    pages = "1805--1819",
    year = "2004"
}

@article{Konoplya:2002ky,
    author = "Konoplya, R. A.",
    title = "{Decay of charged scalar field around a black hole: Quasinormal modes of R-N, R-N-AdS black hole}",
    eprint = "gr-qc/0207028",
    archivePrefix = "arXiv",
    doi = "10.1103/PhysRevD.66.084007",
    journal = "Phys. Rev. D",
    volume = "66",
    pages = "084007",
    year = "2002"
}

\end{document}